\PassOptionsToPackage{unicode}{hyperref}
\PassOptionsToPackage{hyphens}{url}
\documentclass[
  acmsmall,
  screen]{acmart}
\usepackage{amsmath,amsfonts}
\usepackage{xcolor}
\usepackage{color}
\usepackage{fancyvrb}

\DefineVerbatimEnvironment{Highlighting}{Verbatim}{commandchars=\\\{\}}
\newenvironment{Shaded}{}{}

\newcommand{\DataTypeTok}[1]{\textcolor[rgb]{0.56,0.13,0.00}{#1}}

\newcommand{\KeywordTok}[1]{\textcolor[rgb]{0.00,0.44,0.13}{\textbf{#1}}}
\newcommand{\NormalTok}[1]{#1}
\newcommand{\OperatorTok}[1]{\textcolor[rgb]{0.40,0.40,0.40}{#1}}
\newcommand{\OtherTok}[1]{\textcolor[rgb]{0.00,0.44,0.13}{#1}}

\usepackage{longtable,booktabs,array}
\providecommand{\tightlist}{%
  \setlength{\itemsep}{0pt}\setlength{\parskip}{0pt}}
\usepackage{mathtools}
\usepackage{amsmath}
\usepackage{graphicx}
\usepackage{tikz}
\usetikzlibrary{patterns}
\usepackage{pgfplots}
\usepgfplotslibrary{groupplots,units}
\pgfplotsset{compat=1.9}

\newtheorem{example}{Example}

\usepackage[]{natbib}
\usepackage{bookmark}
\hypersetup{
  pdftitle={A formal specification of the jq language},
  pdfkeywords={jq, JSON, semantics},
  hidelinks,
  pdfcreator={LaTeX via pandoc}}

\title{A formal specification of the jq language}

\author{Michael Färber}
\orcid{0000-0003-1634-9525}
\email{michael.faerber@gedenkt.at}


\date{}

\begin{document}
\begin{abstract}
\texttt{jq} is a widely used tool that provides a programming language to manipulate JSON data. However, the jq language is currently only specified by its implementation, making it difficult to reason about its behaviour. To this end, we provide a formal syntax and denotational semantics for a large subset of the jq language. In particular, we give a translation from jq programs to lambda terms. Our most significant contribution is to provide a new way to interpret updates that allows for more performant execution. We implement our semantics in two interpreters and evaluate the performance.
\end{abstract}
\maketitle

UNIX has popularised the concept of \emph{filters} and \emph{pipes} \citep{Ritchie84}: A filter is a program that reads from an input stream and writes to an output stream. Pipes are used to compose filters.

JSON (JavaScript Object Notation) is a widely used data serialisation format \citep{rfc8259}. A JSON value is either null, a boolean, a number, a string, an array of values, or an associative map from strings to values.

\texttt{jq} is a widely used tool that provides a language to define programs called filters and executes them. Where UNIX filters operate on streams of characters, jq filters operate on streams of JSON values. This allows to manipulate JSON data with relatively compact filters. We refer to the program as ``\texttt{jq}'' and to its language as ``the jq language'' or ``jq''. A ``jq interpreter'' is a program that executes jq programs.

The jq language is a dynamically typed, lazily evaluated functional programming language with second-class higher-order functions \citep{jq-description}. The semantics of the jq language are only informally specified, for example in the jq manual \citep{jq-manual}. This leaves a lot of space for interpretation and makes it difficult to find out whether certain behaviour of a jq interpreter is accidental or intended.

We have created denotational semantics (\autoref{sec:semantics}) for the jq language. For jq users, this provides a succinct and unambiguous reference of the language. We cover only the core of jq, i.e.~jq filters with dedicated syntax, whose behaviour is relatively consistent among jq interpreters and stable over time. We do not describe the various named filters for regular expressions, mathematics, formatting, etc. that make up jq's ``standard library'', as this would go beyond the constraints of this article.

Having formal semantics for jq makes it possible to verify the correctness of jq programs and interpreters. For example, our semantics could be used to prove that an optimisation technique in a jq interpreter is correct. Furthermore, our semantics are abstract over the type of processed values. This accommodates the fact that the type of values differs between different jq interpreters and different versions of \texttt{jq}. It also makes it possible to describe jq interpreters that process other kinds of values than JSON.

One of the least specified, yet most fascinating features of jq are \emph{updates}: An update filter, such as \texttt{f\ \textbar{}=\ g}, modifies input data using a filter \texttt{f} that defines which parts of the input to update, and a filter \texttt{g} that defines what the matching input parts should be replaced with. Most jq interpreters perform updates using an approach that we call path-based updates. We found a new approach called \emph{pathless updates} which can be described compactly and unambiguously, eliminates many potential errors, and allows for more performant execution.

We have implemented two jq interpreters: \emph{ujq} is a direct translation of our semantics to Haskell, and \emph{jaq} is an optimised version written in Rust. To test whether a jq interpreter (such as jaq) confirms to our semantics, we can compare its behaviour to ujq.

\autoref{sec:tour} introduces jq by a series of examples that give a glimpse of actual jq syntax and behaviour. \autoref{sec:preliminaries} defines several data types and corresponding lambda terms, such as values, results, and lists. \autoref{sec:syntax} formalises a subset of jq syntax and shows how to transform jq syntax to an intermediate representation (IR). \autoref{sec:semantics} shows how to evaluate jq filters on a given input value. \autoref{sec:updates} presents the traditional path-based and our new pathless approach to executing updates. \autoref{sec:impl} describes and evaluates our two jq interpreters. It shows that our semantics enable correct execution of several large programs written in jq, as well as significantly more performant update execution.

\newcommand{\mathoc}[1]{\mathopen{}#1\mathclose{}}

\newcommand{\jqf }[1]{\mathtt{#1}}
\newcommand{\jqkw}[1]{\mathrel{\jqf{#1}}}
\newcommand{\jqstr}[1]{\text{"#1"}}
\newcommand{\jqas}{\jqkw{as}}
\newcommand{\jqdef }[2]{\mathoc{\jqkw{def} #1\!: #2;\;}}
\newcommand{\jqite }[3]{\mathoc{\jqkw{if} #1 \jqkw{then} #2 \jqkw{else} #3 \jqkw{end}}}
\newcommand{\jqfold}[4]{\mathoc{#1 #2 \jqas #3\; #4}}
\newcommand{\jqreduce }[3]{\jqfold{\jqkw{reduce}} {#1}{#2}{#3}}
\newcommand{\jqforeach}[3]{\jqfold{\jqkw{foreach}}{#1}{#2}{#3}}
\newcommand{\jqlb  }[2]{\mathoc{\jqkw{#1}\, \$#2}}
\newcommand{\jqtc  }[2]{\mathoc{\jqkw{try} #1 \jqkw{catch} #2}}

\newcommand{\gror}{\mathbin{\quad\|\quad}}
\newcommand{\cartesian}{\mathrel{\circ}}
\newcommand{\update}{\mathbin{|{=}}}
\newcommand{\alt}{\mathbin{/\!/}}
\newcommand{\iseq}{\stackrel{?}{=}}
\newcommand{\modulo}{\mathbin{\%}}
\newcommand{\arith}{\odot}
\newcommand{\fold}{\phi}
\newcommand{\floor}[1]{\lfloor#1\rfloor}
\newcommand{\ceil}[1]{\lceil#1\rceil}
\newcommand{\bool}{\operatorname{bool}}
\newcommand{\true}{\operatorname{true}}
\newcommand{\false}{\operatorname{false}}
\newcommand{\length}{\operatorname{length}}
\newcommand{\stream}[1]{\langle#1\rangle}
\newcommand{\obj}[1]{\left\{#1\right\}}
\newcommand{\eval}{\operatorname{eval}}
\newcommand{\sem}[1]{\left[\!\left[#1\right]\!\right]}
\newcommand{\pair}{\operatorname{pair}}
\newcommand{\run}{\operatorname{run}}
\newcommand{\upd}{\operatorname{upd}}
\newcommand{\ok}{\operatorname{ok}}
\newcommand{\err}{\operatorname{err}}
\newcommand{\breakf}{\operatorname{break}}
\newcommand{\zero}{\operatorname{zero}}
\newcommand{\succf}{\operatorname{succ}}
\newcommand{\nateq}{\operatorname{nat\_eq}}
\newcommand{\nil}{\operatorname{nil}}
\newcommand{\cons}{\operatorname{cons}}
\newcommand{\nullf}{\operatorname{null}}
\newcommand{\arr}{\operatorname{arr}}
\newcommand{\objf}{\operatorname{obj}}
\newcommand{\sumf}{\operatorname{sum}}
\newcommand{\trues}{\operatorname{trues}}
\newcommand{\dom}{\operatorname{dom}}
\newcommand{\keys}{\operatorname{keys}}
\newcommand{\bind}{\mathop{>\!\!\!>\!=}\nolimits}
\newcommand{\bindr}{\bind_{\mathcal{R}}}
\newcommand{\bindl}{\bind_{\stream{}}}
\newcommand{\map}{\operatorname{map}}
\newcommand{\mapr}{\map_{\mathcal{R}}}
\newcommand{\mapl}{\map_{\stream{}}}

\newcommand{\boolt}{\mathbb{B}}
\newcommand{\valt}{\mathcal{V}}
\newcommand{\optt}{\mathcal{O}}
\newcommand{\resultt}{\mathcal{R}}
\newcommand{\filtert}{\mathcal{F}}
\newcommand{\valpatht}{\mathcal{V}_{\mathcal{P}}}

\section{Tour of jq}\label{sec:tour}

This goal of this section is to convey an intuition about jq. The official documentation of jq is its user manual \citep{jq-manual}.

jq programs are called \emph{filters}. For now, let us consider a filter to be a function from a value to a (lazy, possibly infinite) stream of values.

The identity filter ``\texttt{.}'' returns a stream containing the input.\footnote{ The filters in this section can be executed on most UNIX shells by \texttt{echo\ \$INPUT\ \textbar{}\ jq\ \$FILTER}, where \texttt{\$INPUT} is the input value in JSON format and \texttt{\$FILTER} is the jq program to be executed. Often, it is convenient to quote the filter; for example, to run the filter ``\texttt{.}'' with the input value \texttt{0}, we can run \texttt{echo\ 0\ \textbar{}\ jq\ \textquotesingle{}.\textquotesingle{}}. In case where the input value does not matter, we can also use \texttt{jq\ -n\ \$FILTER}, which runs the filter with the input value \texttt{null}. We use \texttt{jq} 1.8.1.}

Arithmetic operations, such as addition, subtraction, multiplication, division, and remainder, are available in jq. For example, ``\texttt{.\ +\ 1}'' returns a stream containing the successor of the input. Here, ``\texttt{1}'' is a filter that returns the value \texttt{1} for any input.

The filter ``\texttt{f,\ g}'' concatenates the outputs of the filters \texttt{f} and \texttt{g}. For example, the filter ``\texttt{.,\ .}'' returns a stream containing the input value twice. The filter ``\texttt{f\ \textbar{}\ g}'' maps the filter \texttt{g} over all outputs of the filter \texttt{f}. For example, ``\texttt{(1,\ 2,\ 3)\ \textbar{}\ (.\ +\ 1)}'' returns \texttt{2,\ 3,\ 4}.

The filter ``\texttt{{[}f{]}}'' creates an array from a stream produced by \texttt{f}. For example, the filter ``\texttt{{[}1,\ 2,\ 3{]}}'' concatenates the output of the filters ``\texttt{1}'', ``\texttt{2}'', and ``\texttt{3}'' and puts it into an array, yielding the value \texttt{{[}1,\ 2,\ 3{]}}. The inverse filter ``\texttt{.{[}{]}}'' returns a stream containing the values of an array if the input is an array. For example, running ``\texttt{.{[}{]}}'' on the array \texttt{{[}1,\ 2,\ 3{]}} yields the stream \texttt{1,\ 2,\ 3} consisting of three values. We can combine the two shown filters to map over arrays; for example, when given the input \texttt{{[}1,\ 2,\ 3{]}}, the filter ``\texttt{{[}.{[}{]}\ \textbar{}\ (.\ +\ 1){]}}'' returns a single value \texttt{{[}2,\ 3,\ 4{]}}. The values of an array at indices produced by \texttt{f} are returned by ``\texttt{.{[}f{]}}''. For example, given the input \texttt{{[}1,\ 2,\ 3{]}}, the filter ``\texttt{.{[}0,\ 2,\ 0{]}}'' returns the stream \texttt{1,\ 3,\ 1}.

Case distinctions can be performed with the filter ``\texttt{if\ f\ then\ g\ else\ h\ end}''. For every value \texttt{v} produced by \texttt{f}, this filter returns the output of \texttt{g} if \texttt{v} is true and the output of \texttt{h} otherwise. For example, given the input \texttt{1}, the filter ``\texttt{if\ (.\ \textless{}\ 1,\ .\ ==\ 1,\ .\ \textgreater{}=\ 1)\ then\ .\ else\ {[}{]}\ end}'' returns \texttt{{[}{]},\ 1,\ 1}.

We can define filters by using the syntax ``\texttt{def\ f(x1;\ ...;\ xn):\ g;}'', which defines an filter \texttt{f} taking \texttt{n} arguments by \texttt{g}, where \texttt{g} can refer to \texttt{x1} to \texttt{xn}. For example, jq provides the filter ``\texttt{recurse(f)}'' to calculate fix points, which could be defined by ``\texttt{def\ recurse(f):\ .,\ (f\ \textbar{}\ recurse(f));}''. Using this, we can define a filter to calculate the factorial function, for example.

\begin{example}[Factorial]\label{ex:fac}

Let us define a filter \texttt{fac} that should return \(n!\) for any input number \(n\). We will define \texttt{fac} using the fix point of a filter \texttt{update}. The input and output of \texttt{update} shall be an array \texttt{{[}n,\ acc{]}}, satisfying the invariant that the final output is \texttt{acc} times the factorial of \texttt{n}. The initial value passed to \texttt{update} is the array ``\texttt{{[}.,\ 1{]}}''. We can retrieve \texttt{n} from the array with ``\texttt{.{[}0{]}}'' and \texttt{acc} with ``\texttt{.{[}1{]}}''. We can now define \texttt{update} as ``\texttt{if\ .{[}0{]}\ \textgreater{}\ 1\ then\ {[}.{[}0{]}\ -\ 1,\ .{[}0{]}\ *\ .{[}1{]}{]}\ else\ empty\ end}'', where ``\texttt{empty}'' is a filter that returns an empty stream. Given the input value \texttt{4}, the filter ``\texttt{{[}.,\ 1{]}\ \textbar{}\ recurse(update)}'' returns \texttt{{[}4,\ 1{]},\ {[}3,\ 4{]},\ {[}2,\ 12{]},\ {[}1,\ 24{]}}. We are, however, only interested in the accumulator contained in the last value. So we can write ``\texttt{{[}.,\ 1{]}\ \textbar{}\ last(recurse(update))\ \textbar{}\ .{[}1{]}}'', where ``\texttt{last(f)}'' is a filter that outputs the last output of \texttt{f}. This then yields a single value \texttt{24} as result.

\end{example}

Composition can also be used to bind values to \emph{variables}. The filter ``\texttt{f\ as\ \$x\ \textbar{}\ g}'' performs the following: Given an input value \texttt{i}, for every output \texttt{o} of the filter \texttt{f} applied to \texttt{i}, the filter binds the variable \texttt{\$x} to the value \texttt{o}, making it accessible to \texttt{g}, and yields the output of \texttt{g} applied to the original input value \texttt{i}. For example, the filter ``\texttt{(0,\ 2)\ as\ \$x\ \textbar{}\ ((1,\ 2)\ as\ \$y\ \textbar{}\ (\$x\ +\ \$y))}'' yields the stream \texttt{1,\ 2,\ 3,\ 4}. Note that in this particular case, we could also write this as ``\texttt{(0,\ 2)\ +\ (1,\ 2)}'', because arithmetic operators such as ``\texttt{f\ +\ g}'' take as inputs the Cartesian product of the output of \texttt{f} and \texttt{g}.\footnote{ Haskell users might appreciate the similarity of the two filters to their Haskell analoga ``\texttt{{[}0,\ 2{]}\ \textgreater{}\textgreater{}=\ (\textbackslash{}x\ -\textgreater{}\ {[}1,\ 2{]}\ \textgreater{}\textgreater{}=\ (\textbackslash{}y\ -\textgreater{}\ return\ (x+y)))}'' and ``\texttt{(+)\ \textless{}\$\textgreater{}\ {[}0,\ 2{]}\ \textless{}*\textgreater{}\ {[}1,\ 2{]}}'', which both return \texttt{{[}1,\ 2,\ 3,\ 4{]}}.}

Folding over streams can be done using \texttt{reduce} and \texttt{foreach}: The filter ``\texttt{reduce\ xs\ as\ \$x\ (init;\ f)}'' keeps a state that is initialised with the output of \texttt{init}. For every element \texttt{\$x} yielded by the filter \texttt{xs}, \texttt{reduce} feeds the current state to the filter \texttt{f}, which may reference \texttt{\$x}, then sets the state to the output of \texttt{f}. When all elements of \texttt{xs} have been yielded, \texttt{reduce} returns the current state. For example, the filter ``\texttt{reduce\ .{[}{]}\ as\ \$x\ (0;\ .\ +\ \$x)}'' calculates the sum over all elements of an array. Similarly, ``\texttt{reduce\ .{[}{]}\ as\ \$x\ (0;\ .\ +\ 1)}'' calculates the length of an array. These two filters are called ``\texttt{add}'' and ``\texttt{length}'' in jq, and they allow to calculate the average of an array by ``\texttt{add\ /\ length}''. The filter ``\texttt{foreach\ xs\ as\ \$x\ (init;\ f)}'' is similar to \texttt{reduce}, but also yields all intermediate states, not only the last state. For example, ``\texttt{foreach\ .{[}{]}\ as\ \$x\ (0;\ .\ +\ \$x)}'' yields the cumulative sum over all array elements.

Wherever we can use \texttt{as\ \$x} to bind values to a variable, we can also destructure values using a \emph{pattern}. For example, given the input value \texttt{{[}1,\ {[}2{]},\ 3{]}}, the filter ``\texttt{.\ as\ {[}\$x,\ {[}\$y{]},\ \$z{]}\ \textbar{}\ \$y}'' yields \texttt{2}.

We can halt filter evaluation with \texttt{label}-\texttt{break}: The filter ``\texttt{label\ \$x\ \textbar{}\ f}'' yields all outputs of \texttt{f} until \texttt{f} invokes ``\texttt{break\ \$x}'', at which point \texttt{f} is not evaluated further. For example, the filter ``\texttt{(label\ \$x\ \textbar{}\ 1,\ break\ \$x,\ 2),\ 3}'' yields the stream \texttt{1,\ 3}. It is possible to break out from arguments passed to filters; for example, the filter ``\texttt{label\ \$x\ \textbar{}\ recurse(break\ \$x)}'' returns its input. We can also nest labels; for example, the filter ``\texttt{label\ \$x\ \textbar{}\ 1,\ (label\ \$y\ \textbar{}\ 2,\ break\ \$x,\ 3),\ 4}'' yields \texttt{1,\ 2}. If we replace \texttt{break\ \$x} by \texttt{break\ \$y}, then it yields \texttt{1,\ 2,\ 4}. If we replace \texttt{break\ \$x} by \texttt{empty}, then it yields \texttt{1,\ 2,\ 3,\ 4}.

The operator ``\texttt{\textbar{}=}'' updates values. It has a similar function as lens setters in languages such as Haskell \citep{FosterPP08, FosterGMPS05, PickeringGW17}: Intuitively, the filter ``\texttt{p\ \textbar{}=\ f}'' considers any part \texttt{v} of the input value returned by \texttt{p} and replaces it by the output of \texttt{f} applied to \texttt{v}. For example, when given the input \texttt{{[}1,\ 2,\ 3{]}}, the filter ``\texttt{.{[}{]}\ \textbar{}=\ (.\ +\ 1)}'' yields \texttt{{[}2,\ 3,\ 4{]}}, and the filter ``\texttt{.{[}1{]}\ \textbar{}=\ (.\ +\ 1)}'' yields \texttt{{[}1,\ 3,\ 3{]}}. We can also nest these filters; for example, when given the input \texttt{{[}{[}1,\ 2{]},\ {[}3,\ 4{]}{]}}, the filter ``\texttt{(.{[}{]}\ \textbar{}\ .{[}{]})\ \textbar{}=\ (.\ +\ 1)}'' yields \texttt{{[}{[}2,\ 3{]},\ {[}4,\ 5{]}{]}}. However, not every filter can be used on the left-hand side of an update; for example, ``\texttt{1\ \textbar{}=\ f}'' yields an error because ``\texttt{1}'' does not point to any part of the input value but creates a new value.

\section{Preliminaries}\label{sec:preliminaries}

In this section, we define several data types, such as values, results, and lists, in simply typed lambda calculus.

\subsection{Basic data types}\label{basic-data-types}

\newcommand{\some}{\operatorname{some}}
\newcommand{\none}{\operatorname{none}}
\newcommand{\fst}{\operatorname{fst}}
\newcommand{\snd}{\operatorname{snd}}

We use letters such as \(T\) and \(U\) for type variables. Let us fix some standard encodings for boolean values \(\mathbb{B}\), natural numbers \(\mathbb N\), optional values \(\mathcal{O}\, T\), lists \(\langle T\rangle\), and pairs \((T, U)\): \[
\begin{alignedat}{4}
\operatorname{true}&\!: \mathbb{B}&&\coloneqq \lambda t\, f. t \\
\operatorname{false}&\!: \mathbb{B}&&\coloneqq \lambda t\, f. f \\
\operatorname{succ}&\!: \mathbb N \to \mathbb N &&\coloneqq \lambda n\, s\, z. s\, n \\
\operatorname{zero}&\!: \mathbb N &&\coloneqq \lambda s\, z. z \\
\operatorname{fst}&{}\!: (T, U) \to T &&\coloneqq \lambda p. p\, (\lambda x\, y. x) \\
\operatorname{snd}&{}\!: (T, U) \to T &&\coloneqq \lambda p. p\, (\lambda x\, y. y)
\end{alignedat}
\qquad
\begin{alignedat}{4}
\operatorname{some}&\!: T \to \mathcal{O}\,T &&\coloneqq \lambda x\, s\, n. s\, x \\
\operatorname{none}&\!: \mathcal{O}\,T &&\coloneqq \lambda s\, n. n \\
\operatorname{cons}&\!: T \to \langle T\rangle \to \langle T\rangle &&\coloneqq \lambda h\, t. \lambda c\, n. c\, h\, t \\
\operatorname{nil}&\!: \langle T\rangle &&\coloneqq \lambda c\, n. n \\
\operatorname{pair}&{}\!: T \to U \to (T, U) &&\coloneqq \lambda x\, y\, p. p\, x\, y \\
\quad
\end{alignedat}
\] We write the empty list ``\(\operatorname{nil}\)'' as \(\langle\rangle\) and ``\(\operatorname{cons}\, r_1\, (\operatorname{cons}\, r_2\, ...)\)'' as \(\langle r_1, r_2, ...\rangle\). Because the jq language is evaluated lazily, lists can be infinite.

We assume a combinator \(Y\!: (T \to T) \to T\) for which \(Y\, f = f\, (Y\, f)\) holds. We will use this to define recursive functions; for example, the concatenation of two lists \(l\) and \(r\) is \[l + r \coloneqq Y\, (\lambda f\, l. l\, (\lambda h\, t. \operatorname{cons}\, h\, (f\, t))\, r)\, l,\] which satisfies the property \[l + r = l\, (\lambda h\, t. \operatorname{cons}\, h\, (t + r))\, r.\] For simplicity, we will define recursive functions from here on mostly by equational properties, from which we could easily derive proper definitions using the \(Y\) combinator.

\subsection{Values}\label{sec:values}

While \texttt{jq} operates uniquely on JSON values, we define jq semantics for a general value type \(\mathcal{V}\). Let us start by defining a few types related to values.

A \emph{result} \(\mathcal{R}\, T\) is either OK, an error, or a break. Non-OK results are also called \emph{exceptions}. \begin{align*}
\operatorname{ok}&{}\!: T \to \mathcal{R}\, T \coloneqq \lambda x\, o\, e\, b. o\, x \\
\operatorname{err}&{}\!: \mathcal{V}\to \mathcal{R}\, T \coloneqq \lambda x\, o\, e\, b. e\, x \\
\operatorname{break}&{}\!: \mathbb N \to \mathcal{R}\, T \coloneqq \lambda x\, o\, e\, b. b\, x
\end{align*} A \emph{value-path} \(\mathcal{V}_{\mathcal{P}}= (\mathcal{V},\, \mathcal{O}\, \langle\mathcal{V}\rangle)\) is a pair of a value and an optional list of values. The optional list is called the \emph{path} of the value. The idea behind it is that if the path is present, it represents the location of the value in input data.

We now specify conditions for \(\mathcal{V}\). \autoref{sec:json} gives concrete definitions for JSON values.

For the value type \(\mathcal{V}\), there must be a type of numbers and a type of strings, such that for any number \(n\), \(n\) is a value, and for any string \(s\), \(s\) is a value. Furthermore, for any boolean \(b\), \(b\) is a value. By convention, we write \(v\) for values, \(n\) for numbers, and \(s\) for strings.

The value type must provide the binary arithmetic operations \(\{+, -, \times, \div, \mathbin{\%}\}\) of type \(\mathcal{V}\to \mathcal{V}\to \mathcal{R}\, \mathcal{V}\), as well as a unary negation operation ``\(-\)'' of type \(\mathcal{V}\to \mathcal{R}\, \mathcal{V}\). That means that every arithmetic operation can fail. The value type must also provide Boolean operations \(\{<, \leq, >, \geq, \stackrel{?}{=}, \neq\}\) of type \(\mathcal{V}\to \mathcal{V}\to \mathcal{V}\). Here, \(l \stackrel{?}{=}r\) returns whether \(l\) equals \(r\), and \(l \neq r\) returns its negation.

We assume the existence of several functions:

\begin{itemize}
\tightlist
\item
  \(\operatorname{arr}\!: \langle\mathcal{R}\, \mathcal{V}\rangle \to \mathcal{R}\, \mathcal{V}\) yields an array result from a list of value results.
\item
  \(\operatorname{obj}_0\!: \mathcal{V}\) yields an empty object.
\item
  \(\operatorname{obj}_1\!: \mathcal{V}\to \mathcal{V}\to \mathcal{R}\, \mathcal{V}\) constructs a singleton object from a key and value. (It returns a value result instead of a value because it may fail in case that the provided value is not a valid key, i.e.~a string in JSON.)
\item
  \(\operatorname{bool}\!: \mathcal{V}\to \mathbb{B}\) takes a value and returns a boolean.
\end{itemize}

A \emph{position} \(p\) is defined by the grammar \(p \coloneqq \quad \emptyset \mathbin{\quad\|\quad}v \mathbin{\quad\|\quad}v\!: \mathbin{\quad\|\quad}:v \mathbin{\quad\|\quad}v:v\).\footnote{ We write \(l \coloneqq \quad r_1 \mathbin{\quad\|\quad}\dots \mathbin{\quad\|\quad}r_n\) to say that \(l\) is of shape \(r_i\) for some \(i \leq n\).} The values contained in a position are called \emph{indices}. We assume two operators:

\begin{itemize}
\tightlist
\item
  The \emph{access operator} \(v[p]\) extracts values contained within the value-path \(v\) at position \(p\), yielding a list of value-path results \(\langle\mathcal{R}\, \mathcal{V}_{\mathcal{P}}\rangle\). This operator will be used in \autoref{sec:semantics}.
\item
  The \emph{update operator}\footnote{Although jq's update operator ``\texttt{\textbar{}=}'' resembles the semantic entailment relation ``\(\models\)'', it has a completely different meaning, deriving from the combination of composition ``\texttt{\textbar{}}'' and update ``\texttt{=}''. We do not use semantic entailment ``\(\models\)'' in this paper.} \(v[p] \mathbin{|{=}}f\) replaces those elements \(v' = v[p]\) in the value \(v\) by the output of \(f\, v'\), where \(f\!: \mathcal{V}\to \langle\mathcal{R}\, \mathcal{V}\rangle\). The update operator yields a single value result \(\mathcal{R}\, \mathcal{V}\). This operator will be used in \autoref{sec:updates}.
\end{itemize}

We write \([]\) instead of \([\emptyset]\) and define error-suppressing variants of these operators: If \(v[p]\) yields an error, then \(v[p]?\) yields \(\langle\rangle\) and \(v[p]? \mathbin{|{=}}f\) yields ``\(\operatorname{ok}\, v\)''. Otherwise, \(v[p]? = v[p]\) and \((v[p]? \mathbin{|{=}}f) = (v[p] \mathbin{|{=}}f)\).

\subsection{Composition}\label{composition}

We define three monadic bind operators to describe composition: \begin{alignat*}{7}
&\mathop{>\!\!\!>\!=}\nolimits_{\mathcal{R}}&&{}\!: \mathcal{R}\, T &&\to (T &&\to \mathcal{R}\, U&&) &&\to \mathcal{R}\, U &&\coloneqq \lambda r\, f. r\, (\lambda o. f\, o)\, (\lambda e. r)\, (\lambda b. r) \\
&\mathop{>\!\!\!>\!=}\nolimits_{\langle\rangle}&&{}\!: \langle T\rangle &&\to (T &&\to \langle U\rangle&&) &&\to \langle U\rangle &&\coloneqq \lambda l\, f. l\, (\lambda h\, t. f\, h + (t \mathop{>\!\!\!>\!=}\nolimits_{\langle\rangle}f))\, \operatorname{nil}\\
&\mathop{>\!\!\!>\!=}\nolimits&&{}\!: \langle\mathcal{R}\, T\rangle &&\to (T &&\to \langle\mathcal{R}\, U\rangle&&) &&\to \langle\mathcal{R}\, U\rangle &&\coloneqq \lambda l\, f. l \mathop{>\!\!\!>\!=}\nolimits_{\langle\rangle}(\lambda x. x\, (\lambda o. f\, o)\, (\lambda e. \langle x\rangle)\, (\lambda b. \langle x\rangle))
\end{alignat*} We also define two standard map functions over results and lists: \begin{alignat*}{5}
&\operatorname{map}_{\mathcal{R}}&&{}\!: (T \to U) &&\to \mathcal{R}\, T &&\to \mathcal{R}\, U &&\coloneqq \lambda f\, r. r \mathop{>\!\!\!>\!=}\nolimits_{\mathcal{R}}(\lambda o. \operatorname{ok}\, (f\, o)) \\
&\operatorname{map}_{\langle\rangle}&&{}\!: (T \to U) &&\to \langle T\rangle &&\to \langle U\rangle &&\coloneqq \lambda f\, l. l \mathop{>\!\!\!>\!=}\nolimits_{\langle\rangle}(\lambda x. \langle f\, x\rangle)
\end{alignat*}

\subsection{Implicit conversion}\label{sec:implicit-conversion}

\newcommand{\fromvp}{\operatorname{from\_vp}}
\newcommand{\tovp}{\operatorname{to\_vp}}

We frequently mix values and value-paths. To avoid boilerplate, we assume that in contexts where one of these types is expected, but a value of the other type is given, the value is implicitly converted to match the expected type. A value can be converted to and from a value-path with the following functions: \begin{align*}
\operatorname{to\_vp}&{}\!: \mathcal{V}\to \mathcal{V}_{\mathcal{P}}\coloneqq \lambda v. \operatorname{pair}\, v\, \operatorname{none}\\
\operatorname{from\_vp}&{}\!: \mathcal{V}_{\mathcal{P}}\to \mathcal{V}\coloneqq \operatorname{fst}
\end{align*} This conversion can also take place inside results and lists. For example:

\begin{itemize}
\tightlist
\item
  If we have a value-path \(v_p\!: \mathcal{V}_{\mathcal{P}}\) and apply it to a function that expects a \(\mathcal{V}\), then \(v_p\) is implicitly substituted by ``\(\operatorname{from\_vp}\, v_p\)''.
\item
  If we have a result \(r\!: \mathcal{R}\, \mathcal{V}\) and apply it to a function that expects a \(\mathcal{R}\, \mathcal{V}_{\mathcal{P}}\), then \(r\) is implicitly substituted by ``\(\operatorname{map}_{\mathcal{R}}\, \operatorname{to\_vp}\, r\)''.
\item
  If we have a list \(l\!: \langle\mathcal{R}\, \mathcal{V}\rangle\) and apply it to a function that expects a \(\langle\mathcal{R}\, \mathcal{V}_{\mathcal{P}}\rangle\), then \(l\) is implicitly substituted by ``\(\operatorname{map}_{\langle\rangle}\, (\operatorname{map}_{\mathcal{R}}\, \operatorname{to\_vp})\, l\)''.
\end{itemize}

\section{Syntax}\label{sec:syntax}

We first describe the syntax for a large subset of the jq language in \autoref{sec:hir}. Next, we define a reduced subset of jq syntax called intermediate representation (IR) in \autoref{sec:ir}. We provide a way to translate from jq syntax to IR. We will use IR later to define the semantics in \autoref{sec:semantics}.

\subsection{jq}\label{sec:hir}

We will now present a subset of jq syntax\footnote{ jq provides more syntactical constructs, such as variable arguments, string interpolation, modules, etc. Most of these constructs are light syntactic sugar; for example, the filter \texttt{.a} is equivalent to \texttt{.{[}"a"{]}}. However, because of the sheer amount of such sugar, we omit it here. All these constructs can be transformed into equivalent syntax treated in this text. This is demonstrated by our interpreter jaq, which supports all constructs.} of which we have already seen examples in \autoref{sec:tour}.

A \emph{filter} \(f\) is defined by the grammar \begin{align*}
f &\coloneqq \quad n \mathbin{\quad\|\quad}s \mathbin{\quad\|\quad}. \mathbin{\quad\|\quad}.. \mathbin{\quad\|\quad}\$x \mathbin{\quad\|\quad}(f) \\
  &\mathbin{\quad\|\quad}[] \mathbin{\quad\|\quad}[f] \mathbin{\quad\|\quad}\{f\!: f, \dots, f\!: f\} \\
  &\mathbin{\quad\|\quad}f \star f \mathbin{\quad\|\quad}f \mathrel{\circ}f \mathbin{\quad\|\quad}-f \mathbin{\quad\|\quad}f [p]^? \dots [p]^? \\
  &\mathbin{\quad\|\quad}f \mathrel{\mathtt{as}}P\;|\;f \mathbin{\quad\|\quad}\mathopen{}\mathrel{\mathtt{reduce}} f \mathrel{\mathtt{as}}P\;\; (f;\; f)\mathclose{} \mathbin{\quad\|\quad}\mathopen{}\mathrel{\mathtt{foreach}} f \mathrel{\mathtt{as}}P\;\; (f;\; f;\; f)\mathclose{} \\
  &\mathbin{\quad\|\quad}{\mathopen{}\mathrel{\mathtt{label}}\, \$x\mathclose{}}\;|\;f \mathbin{\quad\|\quad}{\mathopen{}\mathrel{\mathtt{break}}\, \$x\mathclose{}} \\
  &\mathbin{\quad\|\quad}{\mathopen{}\mathrel{\mathtt{if}} f \mathrel{\mathtt{then}} f \mathrel{\mathtt{else}} f \mathrel{\mathtt{end}}\mathclose{}} \mathbin{\quad\|\quad}f? \mathbin{\quad\|\quad}\mathrel{\mathtt{try}} f \mathbin{\quad\|\quad}\mathopen{}\mathrel{\mathtt{try}} f \mathrel{\mathtt{catch}} f\mathclose{} \\
  &\mathbin{\quad\|\quad}{\mathopen{}\mathrel{\mathtt{def}} x\!\!: f;\;\mathclose{} f} \mathbin{\quad\|\quad}\mathopen{}\mathrel{\mathtt{def}} x(x;\; \dots;\; x)\!\!: f;\;\mathclose{} f \\
  &\mathbin{\quad\|\quad}x \mathbin{\quad\|\quad}x(f;\; \dots;\; f)
\end{align*} where:

\begin{itemize}
\tightlist
\item
  \(p\) is a \emph{position} defined by \(p \coloneqq \quad \emptyset \mathbin{\quad\|\quad}f \mathbin{\quad\|\quad}f\!: \mathbin{\quad\|\quad}:f \mathbin{\quad\|\quad}f:f\).
\item
  \(P\) is a \emph{pattern} defined by \(P \coloneqq \quad \$x \mathbin{\quad\|\quad}[P, \dots, P] \mathbin{\quad\|\quad}\{f\!: P, \dots, f\!: P\}\).
\item
  \(x\) is an identifier (such as \(\mathtt{empty}\)).
\item
  \(n\) is a number (such as \(42\) or \(3.14\)).
\item
  \(s\) is a string (such as ``Hello world!'').
\end{itemize}

We use the superscript ``\(?\)'' to denote an optional presence of ``?''; in particular, \(f [p]^? \dots [p]^?\) can be \(f [p]\), \(f [p]?\), \(f [p] [p]\), \(f [p]? [p]\), \(f [p] [p]?\), \(f [p]? [p]?\), \(f [p] [p] [p]\), and so on. The potential instances of the operators \(\star\) and \(\mathrel{\circ}\) are given in \autoref{tab:binops}. All operators \(\star\) and \(\mathrel{\circ}\) are left-associative, except for ``\texttt{\textbar{}}'', ``\texttt{=}'', ``\texttt{\textbar{}=}'', and ``\(\odot\)\texttt{=}''.

We will handle the operators \texttt{reduce} and \texttt{foreach} very similarly; therefore, we introduce \(\phi\) to stand for either \texttt{reduce} or \texttt{foreach}. However, because \texttt{reduce} takes one argument less than \texttt{foreach}, we simply ignore the superfluous argument when handling \texttt{reduce}.

\begin{longtable}[]{@{}lll@{}}
\caption{Binary operators, given in order of increasing precedence. Operators surrounded by parentheses have equal precedence. }\label{tab:binops}\tabularnewline
\toprule\noalign{}
Name & Symbol & Operators \\
\midrule\noalign{}
\endfirsthead
\toprule\noalign{}
Name & Symbol & Operators \\
\midrule\noalign{}
\endhead
\bottomrule\noalign{}
\endlastfoot
Complex & \(\star\) & ``\texttt{\textbar{}}'', ``\texttt{,}'', (``\texttt{=}'', ``\texttt{\textbar{}=}'', ``\(\odot\)\texttt{=}'', ``\texttt{//=}''), ``\texttt{//}'', ``\texttt{or}'', ``\texttt{and}'' \\
Cartesian & \(\mathrel{\circ}\) & (\texttt{==}, \texttt{!=}), (\texttt{\textless{}}, \texttt{\textless{}=}, \texttt{\textgreater{}}, \texttt{\textgreater{}=}), \(\odot\) \\
Arithmetic & \(\odot\) & (\texttt{+}, \texttt{-}), (\texttt{*}, \texttt{/}), \texttt{\%} \\
\end{longtable}

We consider equivalent the following notations\footnote{ Note that in jq, it is invalid syntax to call a nullary filter as \texttt{x()} instead of \texttt{x}, or to define a nullary filter as \texttt{def\ x():\ f;\ g} instead of \texttt{def\ x:\ f;\ g}.}:

\begin{itemize}
\tightlist
\item
  \(f?\) and \(\mathrel{\mathtt{try}} f\)
\item
  \(x\) and \(x()\)
\item
  \(\mathopen{}\mathrel{\mathtt{foreach}} f_x \mathrel{\mathtt{as}}P\;\; (f_y;\; f)\mathclose{}\) and \(\mathopen{}\mathrel{\mathtt{foreach}} f_x \mathrel{\mathtt{as}}P\;\; (f_y;\; f;\; .)\mathclose{}\)
\item
  \(\mathopen{}\mathrel{\mathtt{def}} x\!\!: f;\;\mathclose{} g\) and \(\mathopen{}\mathrel{\mathtt{def}} x()\!\!: f;\;\mathclose{} g\)
\item
  \(\mathrel{\mathtt{if}} f_1 \mathrel{\mathtt{then}} g_1 \mathrel{\mathtt{elif}} f_2 \mathrel{\mathtt{then}} g_2 \dots \mathrel{\mathtt{elif}} f_n \mathrel{\mathtt{then}} g_n \mathrel{\mathtt{else}} h \mathrel{\mathtt{end}}\) and \newline \(\mathopen{}\mathrel{\mathtt{if}} f_1 \mathrel{\mathtt{then}} g_1 \mathrel{\mathtt{else}} (\mathopen{}\mathrel{\mathtt{if}} f_2 \mathrel{\mathtt{then}} g_2 \mathrel{\mathtt{else}} \dots (\mathopen{}\mathrel{\mathtt{if}} f_n \mathrel{\mathtt{then}} g_n \mathrel{\mathtt{else}} h \mathrel{\mathtt{end}}\mathclose{}) \dots \mathrel{\mathtt{end}}\mathclose{}) \mathrel{\mathtt{end}}\mathclose{}\)
\end{itemize}

\subsection{IR}\label{sec:ir}

We now define IR filters and show how to \emph{lower} a jq filter to an IR filter. This allows us later to define semantics for IR in a much less verbose way than for full jq syntax.

An IR filter \(f\) is defined by the grammar \begin{align*}
f &\coloneqq \quad n \mathbin{\quad\|\quad}s \mathbin{\quad\|\quad}. \mathbin{\quad\|\quad}.. \mathbin{\quad\|\quad}\$x \\
  &\mathbin{\quad\|\quad}[] \mathbin{\quad\|\quad}[f] \mathbin{\quad\|\quad}\{\} \mathbin{\quad\|\quad}\{\$x\!: \$x\} \\
  &\mathbin{\quad\|\quad}f \star f \mathbin{\quad\|\quad}\$x \mathrel{\circ}\$x \mathbin{\quad\|\quad}-\$x \mathbin{\quad\|\quad}.[p]^? \\
  &\mathbin{\quad\|\quad}f \mathrel{\mathtt{as}}\$x\;|\;f \mathbin{\quad\|\quad}\mathopen{}\mathrel{\mathtt{reduce}} f \mathrel{\mathtt{as}}\$x\;\; (.;\; f)\mathclose{} \mathbin{\quad\|\quad}\mathopen{}\mathrel{\mathtt{foreach}} f \mathrel{\mathtt{as}}\$x\;\; (.;\; f;\; f)\mathclose{} \\
  &\mathbin{\quad\|\quad}{\mathopen{}\mathrel{\mathtt{if}} \$x \mathrel{\mathtt{then}} f \mathrel{\mathtt{else}} f \mathrel{\mathtt{end}}\mathclose{}} \mathbin{\quad\|\quad}\mathopen{}\mathrel{\mathtt{try}} f \mathrel{\mathtt{catch}} f\mathclose{} \\
  &\mathbin{\quad\|\quad}{\mathopen{}\mathrel{\mathtt{label}}\, \$x\mathclose{}}\;|\;f \mathbin{\quad\|\quad}\mathopen{}\mathrel{\mathtt{break}}\, \$x\mathclose{} \\
  &\mathbin{\quad\|\quad}{\mathopen{}\mathrel{\mathtt{def}} x(x;\; \dots;\; x)\!\!: f;\;\mathclose{} f} \\
  &\mathbin{\quad\|\quad}x(f;\; \dots;\; f)
\end{align*} where \(p\) is a position containing variables instead of filters as indices.

The set of complex operators \(\star\) in IR is reduced compared to full jq syntax --- for example, IR does not include ``\texttt{=}'' and ``\(\odot\)\texttt{=}''. \autoref{tab:op-correspondence} gives an exhaustive list of IR operators and their corresponding operators in jq syntax.

\begin{longtable}[]{@{}llllllllllllllll@{}}
\caption{Binary operators in concrete jq syntax and their corresponding IR operators. }\label{tab:op-correspondence}\tabularnewline
\toprule\noalign{}
\endfirsthead
\endhead
\bottomrule\noalign{}
\endlastfoot
jq & \texttt{\textbar{}} & \texttt{,} & \texttt{\textbar{}=} & \texttt{//} & \texttt{==} & \texttt{!=} & \texttt{\textless{}} & \texttt{\textless{}=} & \texttt{\textgreater{}} & \texttt{\textgreater{}=} & \texttt{+} & \texttt{-} & \texttt{*} & \texttt{/} & \texttt{\%} \\
IR & \(|\) & \(,\) & \(\mathbin{|{=}}\) & \(\mathbin{/\!/}\) & \(\stackrel{?}{=}\) & \(\neq\) & \(<\) & \(\leq\) & \(>\) & \(\geq\) & \(+\) & \(-\) & \(\times\) & \(\div\) & \(\%\) \\
\end{longtable}

Compared to actual jq syntax, IR filters have simpler path operations (such as \(.[p]^?\) versus \(f [p]^? \dots [p]^?\)) and replace certain occurrences of filters by variables (e.g.~\(\$x \mathrel{\circ}\$x\) versus \(f \mathrel{\circ}f\)). This makes their execution easier to describe and understand.

\subsection{Lowering}\label{lowering}

\begin{longtable}[]{@{}ll@{}}
\caption{Lowering of a jq filter \(\phi\) to an IR filter \(\lfloor\phi\rfloor\). }\label{tab:lowering}\tabularnewline
\toprule\noalign{}
\(\varphi\) & \(\lfloor\varphi\rfloor\) \\
\midrule\noalign{}
\endfirsthead
\toprule\noalign{}
\(\varphi\) & \(\lfloor\varphi\rfloor\) \\
\midrule\noalign{}
\endhead
\bottomrule\noalign{}
\endlastfoot
\(n\), \(s\), \(.\), \(..\), \([]\), or \(\{\}\) & \(\varphi\) \\
\(\$x\) or \(\mathopen{}\mathrel{\mathtt{break}}\, \$x\mathclose{}\) & \(\varphi\) \\
\((f)\) & \(\lfloor f\rfloor\) \\
\([f]\) & \([\lfloor f\rfloor]\) \\
\(\{f\!: g\}\) & \(\lfloor f\rfloor \mathrel{\mathtt{as}}\$x'\;|\;\lfloor g\rfloor \mathrel{\mathtt{as}}\$y'\;|\;\{\$x'\!: \$y'\}\) \\
\(\{f_1\!: g_1, \dots, f_n\!: g_n\}\) & \(\lfloor\{f_1\!: g_1\} + \dots + \{f_n\!: g_n\}\rfloor\) \\
\(f\) \texttt{=} \(g\) & \(\lfloor g\rfloor \mathrel{\mathtt{as}}\$x'\;|\;\lfloor f \mathbin{|{=}}\$x'\rfloor\) \\
\(f\) \(\odot\)\texttt{=} \(g\) & \(\lfloor g\rfloor \mathrel{\mathtt{as}}\$x'\;|\;\lfloor f \mathbin{|{=}}(. \odot\$x')\rfloor\) \\
\(f\) \texttt{//=} \(g\) & \(\lfloor f \mathbin{|{=}}. \mathbin{/\!/}g\rfloor\) \\
\(f \mathrel{\mathtt{and}} g\) & \(\lfloor\mathopen{}\mathrel{\mathtt{if}} f \mathrel{\mathtt{then}} (g\;|\;\mathtt{bool}) \mathrel{\mathtt{else}} \mathtt{false} \mathrel{\mathtt{end}}\mathclose{}\rfloor\) \\
\(f \mathrel{\mathtt{or}}  g\) & \(\lfloor\mathopen{}\mathrel{\mathtt{if}} f \mathrel{\mathtt{then}} \mathtt{true} \mathrel{\mathtt{else}} (g\;|\;\mathtt{bool}) \mathrel{\mathtt{end}}\mathclose{}\rfloor\) \\
\(f \star g\) & \(\lfloor f\rfloor \star \lfloor g\rfloor\) \\
\(f \mathrel{\circ}g\) & \(\lfloor f\rfloor \mathrel{\mathtt{as}}\$x'\;|\;\lfloor g\rfloor \mathrel{\mathtt{as}}\$y'\;|\;\$x' \mathrel{\circ}\$y'\) \\
\(-f\) & \(\lfloor f\rfloor \mathrel{\mathtt{as}}\$x'\;|\;-\$x'\) \\
\(f [p_1]^? \dots [p_n]^?\) & \(. \mathrel{\mathtt{as}}\$x'\;|\;\lfloor f\rfloor\;|\;\lfloor[p_1]^?\rfloor_{\$x'}\;|\;\dots\;|\;\lfloor[p_n]^?\rfloor_{\$x'}\) \\
\(f \mathrel{\mathtt{as}}\$x\;|\;g\) & \(\lfloor f\rfloor \mathrel{\mathtt{as}}\$x\;|\;\lfloor g\rfloor\) \\
\(f \mathrel{\mathtt{as}}P\;|\;g\) & \(\lfloor f\rfloor \mathrel{\mathtt{as}}\$x'\;|\;\lfloor\$x' \mathrel{\mathtt{as}}P\;|\;g\rfloor\), \\
\(\$x \mathrel{\mathtt{as}}[P_1, \dots, P_n]\;|\;g\) & \(\lfloor\$x \mathrel{\mathtt{as}}\left\{(0)\!: P_1, \dots, (n-1)\!: P_n\right\}\;|\;g\rfloor\) \\
\(\$x \mathrel{\mathtt{as}}\left\{f_1\!: P_1, \dots\right\}\;|\;g\) & \(\lfloor\$x[f_1] \mathrel{\mathtt{as}}\$x'\;|\;\$x' \mathrel{\mathtt{as}}P_1\;|\;\$x \mathrel{\mathtt{as}}\left\{f_2\!: P_2, \dots\right\}\;|\;g\rfloor\) \\
\(\$x \mathrel{\mathtt{as}}\left\{\right\}\;|\;g\) & \(\lfloor g\rfloor\) \\
\(\mathopen{}\phi f_x \mathrel{\mathtt{as}}\$x\;\; (f_y;\; f;\; g)\mathclose{}\) & \(. \mathrel{\mathtt{as}}\$x'\;|\;\lfloor f_y\rfloor\;|\;\mathopen{}\phi (\$x'\;|\;\lfloor f_x\rfloor) \mathrel{\mathtt{as}}\$x\;\; (.;\; \lfloor f\rfloor;\; \lfloor g\rfloor)\mathclose{}\) \\
\(\mathopen{}\phi f_x \mathrel{\mathtt{as}}P\;\; (f_y;\; f;\; g)\mathclose{}\) & \(\lfloor\mathopen{}\phi (f_x \mathrel{\mathtt{as}}P\;|\;\beta P) \mathrel{\mathtt{as}}\$x'\;\; (f_y;\; \$x' \mathrel{\mathtt{as}}\beta P\;|\;f;\; \$x' \mathrel{\mathtt{as}}\beta P\;|\;g)\mathclose{}\rfloor\) \\
\(\mathopen{}\mathrel{\mathtt{if}} f_x \mathrel{\mathtt{then}} f \mathrel{\mathtt{else}} g \mathrel{\mathtt{end}}\mathclose{}\) & \(\lfloor f_x\rfloor \mathrel{\mathtt{as}}\$x'\;|\;\mathopen{}\mathrel{\mathtt{if}} \$x' \mathrel{\mathtt{then}} \lfloor f\rfloor \mathrel{\mathtt{else}} \lfloor g\rfloor \mathrel{\mathtt{end}}\mathclose{}\) \\
\(f?\) & \(\mathopen{}\mathrel{\mathtt{label}}\, \$x'\mathclose{}\;|\;\mathopen{}\mathrel{\mathtt{try}} \lfloor f\rfloor \mathrel{\mathtt{catch}} (\mathopen{}\mathrel{\mathtt{break}}\, \$x'\mathclose{})\mathclose{}\) \\
\(\mathopen{}\mathrel{\mathtt{try}} f \mathrel{\mathtt{catch}} g\mathclose{}\) & \(\mathopen{}\mathrel{\mathtt{label}}\, \$x'\mathclose{}\;|\;\mathopen{}\mathrel{\mathtt{try}} \lfloor f\rfloor \mathrel{\mathtt{catch}} (\lfloor g\rfloor, \mathopen{}\mathrel{\mathtt{break}}\, \$x'\mathclose{})\mathclose{}\) \\
\(\mathopen{}\mathrel{\mathtt{label}}\, \$x\mathclose{}\;|\;f\) & \(\mathopen{}\mathrel{\mathtt{label}}\, \$x\mathclose{}\;|\;\lfloor f\rfloor\) \\
\(\mathopen{}\mathrel{\mathtt{def}} x\!\!: f;\;\mathclose{} g\) & \(\mathopen{}\mathrel{\mathtt{def}} x\!\!: \lfloor f\rfloor;\;\mathclose{} \lfloor g\rfloor\) \\
\(\mathopen{}\mathrel{\mathtt{def}} x(x_1;\; \dots;\; x_n)\!\!: f;\;\mathclose{} g\) & \(\mathopen{}\mathrel{\mathtt{def}} x(x_1;\; \dots;\; x_n)\!\!: \lfloor f\rfloor;\;\mathclose{} \lfloor g\rfloor\) \\
\(x(f_1;\; \dots;\; f_n)\) & \(x(\lfloor f_1\rfloor;\; \dots;\; \lfloor f_n\rfloor)\) \\
\end{longtable}

\autoref{tab:lowering} shows how to lower a jq filter \(\varphi\) to an equivalent IR filter \(\lfloor\varphi\rfloor\). As a side effect, this provides equational properties that can be used to reason about jq programs. By convention, we write \(\$x'\) to denote a fresh variable. Note that for some complex operators \(\star\), namely ``\texttt{=}'', ``\(\odot\)\texttt{=}'', ``\texttt{//=}'', ``\texttt{and}'', and ``\texttt{or}'', \autoref{tab:lowering} specifies individual lowerings, whereas for the remaining complex operators \(\star\), namely ``\texttt{\textbar{}}'', ``\texttt{,}'', ``\texttt{\textbar{}=}'', and ``\texttt{//}'', \autoref{tab:lowering} specifies a uniform lowering \(\lfloor f \star g\rfloor = \lfloor f\rfloor \star \lfloor g\rfloor\).

We define filters that yield the boolean values as \begin{align*}
\mathtt{true}  &\coloneqq (0 \stackrel{?}{=}0), \\
\mathtt{false} &\coloneqq (0  \neq 0).
\end{align*} The filter ``\(\mathtt{bool} \coloneqq (\mathopen{}\mathrel{\mathtt{if}} . \mathrel{\mathtt{then}} \mathtt{true} \mathrel{\mathtt{else}} \mathtt{false} \mathrel{\mathtt{end}}\mathclose{})\)'' maps its input to its boolean value.

\subsubsection{Compatibility}\label{compatibility}

The lowering in \autoref{tab:lowering} is compatible with the semantics of \texttt{jq}, with one notable exception: In \texttt{jq}, Cartesian operations \(f \mathrel{\circ}g\) are equivalent to \(\lfloor g\rfloor \mathrel{\mathtt{as}}\$y'\;|\;\lfloor f\rfloor \mathrel{\mathtt{as}}\$x'\;|\;\$x \mathrel{\circ}\$y\), whereas we lower them to \(\lfloor f\rfloor \mathrel{\mathtt{as}}\$x'\;|\;\lfloor g\rfloor \mathrel{\mathtt{as}}\$y'\;|\;\$x \mathrel{\circ}\$y\), thus inverting the binding order. Note that the difference only shows when both \(f\) and \(g\) return multiple values. We diverge here from \texttt{jq} to make the lowering of Cartesian operations consistent with that of other operators, such as \(\{f\!: g\}\), where the leftmost filter (\(f\)) is bound first and the rightmost filter (\(g\)) is bound last. That also makes it easier to describe other filters, such as \(\{f_1\!: g_1, \dots, f_n\!: g_n\}\), which we can lower to \(\lfloor\{f_1\!: g_1\} + \dots + \{f_n\!: g_n\}\rfloor\), whereas its lowering assuming \texttt{jq}'s lowering of Cartesian operations would be \[\lfloor\{f_1\!: g_1\}\rfloor \mathrel{\mathtt{as}}\$x'_1\;|\;\dots\;|\;\lfloor\{f_n\!: g_n\}\rfloor \mathrel{\mathtt{as}}\$x'_n\;|\;\$x'_1 + \dots + \$x'_n.\]

\begin{example}

The filter \((0, 2) + (0, 1)\) yields \(\langle 0, 1, 2, 3\rangle\) using our lowering, and \(\langle 0, 2, 1, 3\rangle\) in \texttt{jq}.

\end{example}

\subsubsection{Patterns}\label{patterns}

In the lowering of the folding operators \(\mathopen{}\phi f_x \mathrel{\mathtt{as}}P\;\; (f_y;\; f;\; g)\mathclose{}\) (where \(\phi\) stands for either \(\mathrel{\mathtt{reduce}}\) or \(\mathrel{\mathtt{foreach}}\)), we replace the pattern \(P\) by a variable by ``serialising'' and ``deserialising'' the variables bound by \(P\) with \(\beta P\). Here, \(\beta P\) denotes the sequence of variables bound by \(P\): \[\beta P = \begin{cases}
  \sum_i \beta P_i & \text{if $P = [P_1, \dots, P_n]$ or $P = \{f_1\!: P_1, \dots, f_n\!: P_n\}$} \\
  [\$x] & \text{if $P = \$x$}
\end{cases}\] (We used \(\sum_i x_i = x_1 + \dots + x_n\) and \([x_1, \dots, x_n] + [y_1, \dots, y_m] = [x_1, \dots, x_n, y_1, \dots, y_m]\).) In particular, we exploit the property that \(f \mathrel{\mathtt{as}}P\;|\;g\) can be rewritten to \[ f \mathrel{\mathtt{as}}P\;|\;\beta P \mathrel{\mathtt{as}}\$x'\;|\;\$x' \mathrel{\mathtt{as}}\beta P\;|\;g, \] because \(\beta P\) can be interpreted both as pattern and as filter.

\begin{example}

Consider the filter \(\varphi \equiv f \mathrel{\mathtt{as}}[\$x, [\$y], \$z]\;|\;g\). This filter destructures all outputs of \(f\) that are of the shape \([x, [y, \dots], z, \dots]\) and binds the values \(x\), \(y\), and \(z\) to the respective variables. Here, \(\varphi\) uses the pattern \(P = [\$x, [\$y], \$z]\) for which \(\beta P = [\$x, \$y, \$z]\). It holds that \(\varphi\) is equivalent to \[ f \mathrel{\mathtt{as}}[\$x, [\$y], \$z]\;|\;[\$x, \$y, \$z] \mathrel{\mathtt{as}}\$x'\;|\;\$x' \mathrel{\mathtt{as}}[\$x, \$y, \$z]\;|\;g. \] Here, we first used \(\beta P\) as filter (\([\$x, \$y, \$z] \mathrel{\mathtt{as}}\$x'\;|\;\dots\)) to ``serialise'' the pattern variables to an array, then as pattern (\(\$x' \mathrel{\mathtt{as}}[\$x, \$y, \$z]\;|\;\dots\)) to ``deserialise'' the array to retrieve the pattern variables.

\end{example}

\subsubsection{Positions}\label{positions}

\begin{longtable}[]{@{}ll@{}}
\caption{Lowering of a position \([p]^?\) with input \(\$x\) to an IR filter. }\label{tab:lower-path}\tabularnewline
\toprule\noalign{}
\([p]  ^?\) & \(\lfloor[p]^?\rfloor_{\$x}\) \\
\midrule\noalign{}
\endfirsthead
\toprule\noalign{}
\([p]  ^?\) & \(\lfloor[p]^?\rfloor_{\$x}\) \\
\midrule\noalign{}
\endhead
\bottomrule\noalign{}
\endlastfoot
\([   ]^?\) & \(.[]^?\) \\
\([f  ]^?\) & \((\$x\;|\;\lfloor f\rfloor) \mathrel{\mathtt{as}}\$y'\;|\;.[\$y']^?\) \\
\([f\!: ]^?\) & \((\$x\;|\;\lfloor f\rfloor) \mathrel{\mathtt{as}}\$y'\;|\;.[\$y':]^?\) \\
\([ :f]^?\) & \((\$x\;|\;\lfloor f\rfloor) \mathrel{\mathtt{as}}\$y'\;|\;.[:\$y']^?\) \\
\([f:g]^?\) & \((\$x\;|\;\lfloor f\rfloor) \mathrel{\mathtt{as}}\$y'\;|\;(\$x\;|\;\lfloor g\rfloor) \mathrel{\mathtt{as}}\$z'\;|\;.[\$y':\$z']^?\) \\
\end{longtable}

\autoref{tab:lower-path} shows how to lower a position \([p]^?\) to IR filters. Like in \autoref{sec:hir}, the meaning of superscript ``\(?\)'' is an optional presence of ``\(?\)''. In the lowering of \(f [p_1]^? \dots [p_n]^?\) in \autoref{tab:lowering}, if \([p_i]\) in the first column is directly followed by ``?'', then \(\lfloor[p_i]^?\rfloor_{\$x}\) in the second column stands for \(\lfloor[p_i] ?\rfloor_{\$x}\), otherwise for \(\lfloor[p_i]  \rfloor_{\$x}\). Similarly, in \autoref{tab:lower-path}, if \([p]\) in the first column is followed by ``\(?\)'', then all occurrences of superscript ``?'' in the second column stand for ``?'', otherwise for nothing.

\begin{example}

The jq filter \((.[]?[])\) is lowered to \((. \mathrel{\mathtt{as}}\$x'\;|\;.\;|\;.[]?\;|\;.[])\). Semantically, we will see that this is equivalent to \((.[]?\;|\;.[])\).

\end{example}

\begin{example}

The jq filter \(\mu \equiv .[0]\) is lowered to \(\lfloor\mu \rfloor\equiv (. \mathrel{\mathtt{as}}\$x\;|\;.\;|\;(\$x\;|\;0) \mathrel{\mathtt{as}}\$y\;|\;.[\$y])\). Semantically, we will see that \(\lfloor\mu\rfloor\) is equivalent to \((0 \mathrel{\mathtt{as}}\$y\;|\;.[\$y])\). The jq filter \(\varphi \equiv ([3]\;|\;.[0] = (\mathtt{length}, 2))\) is lowered to the IR filter \(\lfloor\varphi \rfloor\equiv ([3]\;|\;(\mathtt{length}, 2) \mathrel{\mathtt{as}}\$z\;|\;\lfloor\mu \rfloor\mathbin{|{=}}\$z)\). In \autoref{sec:semantics}, we will see that its output is \(\langle[1], [2]\rangle\).

\end{example}

\subsection{Hygiene}\label{sec:hygiene}

For hygienic reasons, we require that labels are disjoint from variables and that filters with the same name represent filters with equal arity. This can be easily ensured by renaming labels \(\$x\) to \(\$l_x\), variables \(\$x\) to \(\$v_x\), and identifiers \(x\) of arity \(n\) to \(x^n\). For brevity, we do not perform this renaming in most examples.

\begin{example}

Consider the filter \(\mathopen{}\mathrel{\mathtt{label}}\, \$x\mathclose{}\;|\;. \mathrel{\mathtt{as}}\$x\;|\;\$x + \$x, \mathopen{}\mathrel{\mathtt{break}}\, \$x\mathclose{}\). Here, we have to rename to ensure that labels and variables are disjoint, yielding e.g. \(\mathopen{}\mathrel{\mathtt{label}}\, \$l_x\mathclose{}\;|\;. \mathrel{\mathtt{as}}\$v_x\;|\;\$v_x + \$v_x, \mathopen{}\mathrel{\mathtt{break}}\, \$l_x\mathclose{}\).

\end{example}

\begin{example}

Consider the filter \(\mathopen{}\mathrel{\mathtt{def}} f(g)\!\!: g;\;\mathclose{} \mathopen{}\mathrel{\mathtt{def}} f\!\!: .;\;\mathclose{} f(f)\), which uses two filters called ``\(f\)'' with different arity. Here, we have to rename identifiers to prevent shadowing issues in the semantics, yielding e.g. \(\mathopen{}\mathrel{\mathtt{def}} f^1(g^0)\!\!: g^0;\;\mathclose{} \mathopen{}\mathrel{\mathtt{def}} f^0\!\!: .;\;\mathclose{} f^1(f^0)\).

\end{example}

\subsection{Examples}\label{examples}

\begin{example}

Consider the following jq program: \begin{align*}
&\mathopen{}\mathrel{\mathtt{def}} \mathtt{empty}\!\!: (\{\}[]) \mathrel{\mathtt{as}}\$x\;|\;.;\;\mathclose{} \\
&\mathopen{}\mathrel{\mathtt{def}} \mathtt{select}(f)\!\!: \mathopen{}\mathrel{\mathtt{if}} f \mathrel{\mathtt{then}} . \mathrel{\mathtt{else}} \mathtt{empty} \mathrel{\mathtt{end}}\mathclose{};\;\mathclose{} \\
&\mathopen{}\mathrel{\mathtt{def}} \mathtt{negative}\!\!: . < 0;\;\mathclose{} \\
&.[]\;|\;\mathtt{select}(\mathtt{negative})
\end{align*} When given an array as an input, it yields those elements of the array that are smaller than \(0\).\footnote{ The filter \texttt{empty} returns an empty stream. We might be tempted to define it as \texttt{\{\}{[}{]}}, which constructs an empty object, then returns its contained values, which corresponds to an empty stream as well. However, such a definition relies on the temporary construction of new values (such as the empty object here), which is not admissible on the left-hand side of updates (see \autoref{sec:updates}). To ensure that \texttt{empty} can be employed also as a path expression, we define it in this complicated manner.} Lowering this to IR yields \begin{align*}
&\mathopen{}\mathrel{\mathtt{def}} \mathtt{empty}\!\!: (. \mathrel{\mathtt{as}}\$x'\;|\;\{\}\;|\;.[]) \mathrel{\mathtt{as}}\$x\;|\;.;\;\mathclose{} \\
&\mathopen{}\mathrel{\mathtt{def}} \mathtt{select}(f)\!\!: f \mathrel{\mathtt{as}}\$x'\;|\;\mathopen{}\mathrel{\mathtt{if}} \$x' \mathrel{\mathtt{then}} . \mathrel{\mathtt{else}} \mathtt{empty} \mathrel{\mathtt{end}}\mathclose{};\;\mathclose{} \\
&\mathopen{}\mathrel{\mathtt{def}} \mathtt{negative}\!\!: . \mathrel{\mathtt{as}}\$x'\;|\;0 \mathrel{\mathtt{as}}\$y'\;|\;\$x' < \$y';\;\mathclose{} \\
&. \mathrel{\mathtt{as}}\$x'\;|\;.\;|\;.[]\;|\;\mathtt{select}(\mathtt{negative})
\end{align*}

\end{example}

\begin{example}

Consider the jq filter \(\mathopen{}\mathrel{\mathtt{def}} \mathtt{recurse}(f)\!\!: ., (f\;|\;\mathtt{recurse}(f));\;\mathclose{} \mathtt{recurse}(. + 1)\), which returns the infinite stream of output values \(n, n+1, \dots\) when provided with an input number \(n\). Lowering this to IR yields \(\mathopen{}\mathrel{\mathtt{def}} \mathtt{recurse}(f)\!\!: ., (f\;|\;\mathtt{recurse}(f));\;\mathclose{} \mathtt{recurse}(. \mathrel{\mathtt{as}}\$x'\;|\;1 \mathrel{\mathtt{as}}\$y'\;|\;\$x' + \$y')\).

\end{example}

\section{Evaluation Semantics}\label{sec:semantics}

In this section, we will show how to transform --- or compile --- an IR filter \(\varphi\) to a lambda term \(\left[\!\left[\varphi\right]\!\right]\) of type \(\mathcal{F}\). We will then define a function ``\(\operatorname{eval}\, \left[\!\left[\varphi\right]\!\right]\, v\)'' that returns the list of value-path results that the filter \(\varphi\) outputs when given the value-path \(v\!: \mathcal{V}_{\mathcal{P}}\) as input. The evaluation strategy is call-by-name.

\subsection{Compilation}\label{compilation}

A compiled filter \(\mathcal{F}\) is a pair of a \emph{run} and an \emph{update} function. We can obtain the run and update functions from a compiled filter as follows: \begin{alignat*}{3}
\operatorname{run}&{}\!: \mathcal{F}\to (\mathcal{V}_{\mathcal{P}}\to \langle\mathcal{R}\, \mathcal{V}_{\mathcal{P}}\rangle) &&\coloneqq \operatorname{fst}\\
\operatorname{upd}&{}\!: \mathcal{F}\to (\mathcal{V}\to \langle\mathcal{R}\, \mathcal{V}\rangle) \to \mathcal{V}\to \langle\mathcal{R}\, \mathcal{V}\rangle &&\coloneqq \operatorname{snd}
\end{alignat*} The lambda term \(\left[\!\left[\varphi \right]\!\right]\!: \mathcal{F}\) corresponding to an IR filter \(\varphi\) has the shape \[\left[\!\left[\varphi \right]\!\right]= \operatorname{pair}\, (\lambda v. t_r)\, (\lambda \sigma\, v. t_u)\] for some terms \(t_r\) (run function) and \(t_u\) (update function). For a given \(\varphi\), we can obtain \(t_r\) by ``\(\operatorname{run}\, \left[\!\left[\varphi\right]\!\right]\, v\)'' and \(t_u\) by ``\(\operatorname{upd}\, \left[\!\left[\varphi\right]\!\right]\, \sigma\, v\)''. For conciseness, we write ``\(\operatorname{run}\, \left[\!\left[\varphi\right]\!\right]\, v\)'' to define \(t_r\) and ``\(\operatorname{upd}\, \left[\!\left[\varphi\right]\!\right]\, \sigma\, v\)'' to define \(t_u\).

\begin{example}[Identity filter compilation]

For the identity filter ``\(.\)'', we have \[ \left[\!\left[.\right]\!\right] = \operatorname{pair}\,
(\lambda v. \langle\operatorname{ok}v\rangle)\,
(\lambda \sigma\, v. \sigma\, v), \] where \(t_r = \langle\operatorname{ok}v\rangle\) was obtained from \autoref{tab:eval-semantics} and \(t_u = \sigma\, v\) was obtained from \autoref{tab:update-semantics}.

\end{example}

\newcommand{\fresh}{\kappa}

The lambda term \(\left[\!\left[\varphi\right]\!\right]\) obtained from a filter \(\varphi\) contains the free variable \(\kappa\). This variable is used to generate fresh labels for the execution of \(\mathopen{}\mathrel{\mathtt{label}}\, \$x\mathclose{}\;|\;f\), see \autoref{ex:labels}. We can then run a filter using the following function: \[\operatorname{eval}\!: \mathcal{F}\to \mathcal{V}\to \langle\mathcal{R}\, \mathcal{V}\rangle \coloneqq \lambda \varphi. (\lambda \kappa. \operatorname{run}\, \varphi)\, \operatorname{zero}\] This initially binds \(\kappa\) to zero, yielding a closed term iff \(\varphi\) is \emph{wellformed}; see \autoref{sec:wf}.

\newcommand{\reducef }{\operatorname{reduce }}
\newcommand{\foreachf}{\operatorname{foreach}}
\newcommand{\untilf}{\operatorname{until}}
\newcommand{\labelf}{\operatorname{label}}

\begin{longtable}[]{@{}ll@{}}
\caption{Evaluation semantics. }\label{tab:eval-semantics}\tabularnewline
\toprule\noalign{}
\(\varphi\) & \(\operatorname{run}\, \left[\!\left[\varphi\right]\!\right]\, v\) \\
\midrule\noalign{}
\endfirsthead
\toprule\noalign{}
\(\varphi\) & \(\operatorname{run}\, \left[\!\left[\varphi\right]\!\right]\, v\) \\
\midrule\noalign{}
\endhead
\bottomrule\noalign{}
\endlastfoot
\(n\) or \(s\) & \(\langle\operatorname{ok}\varphi\rangle\) \\
\(.\) & \(\langle\operatorname{ok}v\rangle\) \\
\(..\) & \(\operatorname{run}\, \left[\!\left[\mathopen{}\mathrel{\mathtt{def}} r\!\!: ., (.[]?\;|\;r);\;\mathclose{} r\right]\!\right]\, v\) \\
\(\$x\) & \(\langle\operatorname{ok}\$x\rangle\) \\
\([  ]\) & \(\langle\operatorname{arr}\, \langle\rangle\rangle\) \\
\(\{\}\) & \(\langle\operatorname{ok}\operatorname{obj}_0\rangle\) \\
\([f ]\) & \(\langle\operatorname{arr}\, (\operatorname{run}\, \left[\!\left[f\right]\!\right]\, v)\rangle\) \\
\(\{\$x\!: \$y\}\) & \(\langle\operatorname{obj}_1\, \$x\, \$y\rangle\) \\
\(\$x \odot\$y\) & \(\langle\$x \odot\$y\rangle\) \\
\(\$x \mathrel{\circ}\$y\) & \(\langle\operatorname{ok}{(\$x \mathrel{\circ}\$y)}\rangle\) \\
\(-\$x\) & \(\langle- \$x\rangle\) \\
\(f, g\) & \(\operatorname{run}\, \left[\!\left[f\right]\!\right]\, v + \operatorname{run}\, \left[\!\left[g\right]\!\right]\, v\) \\
\(f\;|\;g\) & \(\operatorname{run}\, \left[\!\left[f\right]\!\right]\, v \mathop{>\!\!\!>\!=}\nolimits\operatorname{run}\, \left[\!\left[g\right]\!\right]\) \\
\(f \mathbin{/\!/}g\) & \((\operatorname{run}\, \left[\!\left[f\right]\!\right]\, v \mathop{>\!\!\!>\!=}\nolimits\operatorname{trues})\, (\lambda h\, t. \langle h\rangle + t)\, (\operatorname{run}\, \left[\!\left[g\right]\!\right]\, v)\) \\
\(f \mathrel{\mathtt{as}}\$x\;|\;g\) & \(\operatorname{run}\, \left[\!\left[f\right]\!\right]\, v \mathop{>\!\!\!>\!=}\nolimits(\lambda \$x. \operatorname{run}\, \left[\!\left[g\right]\!\right]\, v)\) \\
\(\mathopen{}\mathrel{\mathtt{try}} f \mathrel{\mathtt{catch}} g\mathclose{}\) & \(\operatorname{run}\, \left[\!\left[f\right]\!\right]\, v \mathop{>\!\!\!>\!=}\nolimits_L \lambda r. r\, (\lambda o. \langle r\rangle)\, (\operatorname{run}\, \left[\!\left[g\right]\!\right])\, (\lambda b. \langle r\rangle)\) \\
\(\mathopen{}\mathrel{\mathtt{label}}\, \$x\mathclose{}\;|\;f\) & \(\operatorname{label}\kappa\, ((\lambda \$x\, \kappa. \operatorname{run}\, \left[\!\left[f\right]\!\right]\, v)\, \kappa\, (\operatorname{succ}\, \kappa))\) \\
\(\mathopen{}\mathrel{\mathtt{break}}\, \$x\mathclose{}\) & \(\langle\operatorname{break}\, \$x\rangle\) \\
\(\mathopen{}\mathrel{\mathtt{if}} \$x \mathrel{\mathtt{then}} f \mathrel{\mathtt{else}} g \mathrel{\mathtt{end}}\mathclose{}\) & \(\operatorname{run}\, ((\operatorname{bool}\, \$x)\, \left[\!\left[f\right]\!\right]\, \left[\!\left[g\right]\!\right])\, v\) \\
\(.[p]^?\) & \(v[p]^?\) \\
\(\mathopen{}\mathrel{\mathtt{reduce}} f_x \mathrel{\mathtt{as}}\$x\;\; (.;\; f   )\mathclose{}\) & \(\operatorname{reduce }\, (\lambda \$x. \operatorname{run}\, \left[\!\left[f\right]\!\right])\, (\operatorname{run}\, \left[\!\left[f_x\right]\!\right]\, v)\, v\) \\
\(\mathopen{}\mathrel{\mathtt{foreach}} f_x \mathrel{\mathtt{as}}\$x\;\; (.;\; f;\; g)\mathclose{}\) & \(\operatorname{foreach}\, (\lambda \$x. \operatorname{run}\, \left[\!\left[f\right]\!\right])\, (\lambda \$x. \operatorname{run}\, \left[\!\left[g\right]\!\right])\, (\operatorname{run}\, \left[\!\left[f_x\right]\!\right]\, v)\, v\) \\
\(\mathopen{}\mathrel{\mathtt{def}} x(x_1;\; ...;\; x_n)\!\!: f;\;\mathclose{} g\) & \((\lambda x. \operatorname{run}\, \left[\!\left[g\right]\!\right]\, v) (Y\, (\lambda x\, x_1\, ...\, x_n. \left[\!\left[f\right]\!\right]))\) \\
\(x(f_1;\; ...;\; f_n)\) & \(\operatorname{run}\, (x\, \left[\!\left[f_1\right]\!\right]\, ...\, \left[\!\left[f_n\right]\!\right])\, v\) \\
\(f \mathbin{|{=}}g\) & \(\operatorname{upd}\, \left[\!\left[f\right]\!\right]\, (\operatorname{run}\, \left[\!\left[g\right]\!\right])\, v\) \\
\end{longtable}

The evaluation semantics are given in \autoref{tab:eval-semantics}. This is the very heart of jq. Let us discuss its different cases:

\begin{itemize}
\tightlist
\item
  \(n\) or \(s\): Returns the value corresponding to the number \(n\) or string \(s\).
\item
  ``\(.\)'': Returns the input value. This is the identity filter.
\item
  ``\(..\)'': Returns the input value and all values recursively contained within it.
\item
  \(\$x\): Returns the value currently bound to the variable \(\$x\). Wellformedness of the filter (see \autoref{sec:wf}) ensures that whenever we evaluate a variable, it must have been substituted by a concrete value, for example by a surrounding call to ``\(f \mathrel{\mathtt{as}}\$x\;|\;g\)''.
\item
  \([]\) or \(\{\}\): Creates an empty array or object.
\item
  \([f]\): Creates an array from the output of \(f\), using the ``\(\operatorname{arr}\)'' function defined in \autoref{sec:values}.
\item
  \(\{\$x\!: \$y\}\): Creates an object from the values bound to \(\$x\) and \(\$y\), using the ``\(\operatorname{obj}_1\)'' function defined in \autoref{sec:values}.
\item
  \(\$x \odot\$y\): Returns the output of an arithmetic operation ``\(\odot\)'' (any of \(+\), \(-\), \(\times\), \(\div\), and \(\%\), as given in \autoref{tab:op-correspondence}) on the values bound to \(\$x\) and \(\$y\).
\item
  \(\$x \mathrel{\circ}\$y\): Returns the output of a Boolean operation ``\(\mathrel{\circ}\)'' (any of \(\stackrel{?}{=}\), \(\neq\), \(<\), \(\leq\), \(>\), \(\geq\), as given in \autoref{tab:op-correspondence}) on the values bound to \(\$x\) and \(\$y\). Because we assumed that Boolean operations return \(\mathcal{V}\) and are thus infallible (unlike the arithmetic operations \(\odot\), which return \(\mathcal{R}\, \mathcal{V}\)), we have to wrap their outputs in an ``\(\operatorname{ok}\)''.
\item
  \(f, g\): Concatenates the outputs of \(f\) and \(g\), both applied to the same input.
\item
  \(f\;|\;g\): Composes \(f\) and \(g\), returning the outputs of \(g\) applied to all outputs of \(f\).
\item
  \(f \mathbin{/\!/}g\): Let \(l\) be the outputs of \(f\) whose boolean values are not false. This filter returns \(l\) if \(l\) is not empty, else the outputs of \(g\). Here, we use a function ``\(\operatorname{trues}\)'' that returns its input if its boolean value is true. \[\operatorname{trues}\!: \mathcal{V}_{\mathcal{P}}\to \langle\mathcal{R}\, \mathcal{V}_{\mathcal{P}}\rangle \coloneqq \lambda x. (\operatorname{bool}\, x)\, \langle\operatorname{ok}\, x\rangle\, \langle\rangle\]
\item
  \(f \mathrel{\mathtt{as}}\$x\;|\;g\): For every output of \(f\), binds it to the variable \(\$x\) and returns the output of \(g\), where \(g\) may reference \(\$x\). Unlike \(f\;|\;g\), this runs \(g\) with the original input value instead of an output of \(f\). We can show that the evaluation of \(f\;|\;g\) is equivalent to that of \(f \mathrel{\mathtt{as}}\$x'\;|\;\$x'\;|\;g\), where \(\$x'\) is a fresh variable. Therefore, we could be tempted to lower \(f\;|\;g\) to \(\lfloor f\rfloor \mathrel{\mathtt{as}}\$x'\;|\;\$x'\;|\;\lfloor g\rfloor\) in \autoref{tab:lowering}, in order to further simplify IR and thus the semantics. However, we cannot do this because we will see in \autoref{sec:updates} that this equivalence does \emph{not} hold for updates; that is, \((f\;|\;g) \mathbin{|{=}}\sigma\) is \emph{not} equal to \((f \mathrel{\mathtt{as}}\$x'\;|\;\$x'\;|\;g) \mathbin{|{=}}\sigma\).
\item
  \(\mathopen{}\mathrel{\mathtt{try}} f \mathrel{\mathtt{catch}} g\mathclose{}\): Replaces all outputs of \(f\) that equal ``\(\operatorname{err}\, e\)'' for some \(e\) by the output of \(g\) on the input \(e\). At first sight, this seems to diverge from jq, which aborts the evaluation of \(f\) after the first error. However, because lowering to IR replaces \(\mathrel{\mathtt{try}} f \mathrel{\mathtt{catch}} g\) with \(\mathopen{}\mathrel{\mathtt{label}}\, \$x'\mathclose{}\;|\;\mathopen{}\mathrel{\mathtt{try}} f \mathrel{\mathtt{catch}} (g, \mathopen{}\mathrel{\mathtt{break}}\, \$x'\mathclose{})\mathclose{}\) (see \autoref{tab:lowering}), the overall behaviour described here corresponds to jq after all.
\item
  \(\mathopen{}\mathrel{\mathtt{label}}\, \$x\mathclose{}\;|\;f\): Returns all values yielded by \(f\) until \(f\) yields an exception ``\(\operatorname{break}\, \$x\)''. This uses a function ``\(\operatorname{label}\)'' that takes a label \(\kappa\) and a list \(l\) of value results, returning the longest prefix of \(l\) that does not contain \(\operatorname{break}\, \kappa\): \begin{alignat*}{3}
  \operatorname{label}&{}\!: \mathbb N \to \langle\mathcal{R}\, T\rangle \to \langle\mathcal{R}\, T\rangle &\coloneqq{}&
  \lambda \kappa. \operatorname{until}\, (\lambda r. r\, (\lambda o. \operatorname{false})\, (\lambda e. \operatorname{false})\, (\lambda b. \operatorname{nat\_eq}\, \kappa\, b)) \\
  \operatorname{until}&{}\!: (T \to \mathbb B) \to \langle T\rangle \to \langle T\rangle &{}\coloneqq{}&
  \lambda p\, l. l\, (\lambda h\, t. p\, h\, \langle\rangle\, (\langle h\rangle + \operatorname{until}\, p\, t))
  \end{alignat*} Here, ``\(\operatorname{until}\, p\, l\)'' returns the longest prefix of \(l\) for which every element satisfies \(p\), and \(\operatorname{nat\_eq}\!: \mathbb N \to \mathbb N \to \mathbb{B}\) returns \(\operatorname{true}\) if two natural numbers are equal, else \(\operatorname{false}\).
\item
  \(\mathopen{}\mathrel{\mathtt{break}}\, \$x\mathclose{}\): Returns a value result ``\(\operatorname{break}\, \$x\)''. Similarly to the evaluation of variables \(\$x\) described above, wellformedness of the filter (see \autoref{sec:wf}) ensures that the returned value \(\operatorname{break}\, \$x\) will be eventually handled by a corresponding filter \(\mathopen{}\mathrel{\mathtt{label}}\, \$x\mathclose{}\;|\;f\). That means that \(\operatorname{eval}\left[\!\left[\varphi\right]\!\right]\) for a wellformed filter \(\varphi\) can only yield values and errors, but never a break result.
\item
  \(\mathopen{}\mathrel{\mathtt{if}} \$x \mathrel{\mathtt{then}} f \mathrel{\mathtt{else}} g \mathrel{\mathtt{end}}\mathclose{}\): Returns the output of \(f\) if \(\$x\) is bound to a ``true'' value (neither null nor false for JSON, see \autoref{sec:json}), else returns the output of \(g\).
\item
  \(.[p]^?\): Accesses parts of the input value; see \autoref{sec:values}. Wellformedness ensures that the indices contained in \(p\) have been substituted by values.
\item
  \(\mathopen{}\phi f_x \mathrel{\mathtt{as}}\$x\;\; (.;\; f;\; g)\mathclose{}\): Folds \(f\) over the values returned by \(f_x\), starting with the current input as accumulator. The current accumulator value is provided to \(f\) as input value and \(f\) can access the current value of \(f_x\) by \(\$x\). If \(\phi=  \mathtt{reduce}\), this returns only the final values of the accumulator, whereas if \(\phi= \mathtt{foreach}\), this returns also the intermediate values of the accumulator. We will further explain this and define the functions ``\(\operatorname{reduce }f\,     l\, v\)'' and ``\(\operatorname{foreach}f\, g\, l\, v\)'' in \autoref{sec:folding}.
\item
  \(\mathopen{}\mathrel{\mathtt{def}} x(x_1;\; ...;\; x_n)\!\!: f;\;\mathclose{} g\): Binds the \(n\)-ary filter \(x\) in \(g\). The definition of \(x\), namely \(f\), may refer to any of the arguments \(x_i\) as well as to \(x\) itself. In other words, filters can be defined recursively, which is why we use the \(Y\) combinator here. \autoref{ex:recursion} shows how a recursive call is evaluated.
\item
  \(x(f_1;\; ...;\; f_n)\): Calls an \(n\)-ary filter \(x\). This also handles the case of calling nullary filters such as \(\mathtt{empty}\).
\item
  \(f \mathbin{|{=}}g\): Updates the input at positions returned by \(f\) by \(g\). We will discuss this in \autoref{sec:updates}.
\end{itemize}

\subsection{Examples}\label{examples-1}

\begin{example}[Recursion]\label{ex:recursion}

Consider the following IR filter \(\varphi\): \[\mathopen{}\mathrel{\mathtt{def}} \mathtt{repeat}\!\!: ., \mathtt{repeat};\;\mathclose{} \mathtt{repeat}\] This filter repeatedly outputs its input; for example, given the input \(v = 1\), it returns \(\langle\operatorname{ok}1, \operatorname{ok}1, \operatorname{ok}1, ...\rangle\). First, let us compile a part of our filter, namely \[\rho = \left[\!\left[., \mathtt{repeat}\right]\!\right] =^{\left[\!\left[\cdot\right]\!\right]} \operatorname{pair}\, (\lambda v. \langle\operatorname{ok}v\rangle + \operatorname{run}\, \mathtt{repeat}\, v)\, (...).\] Here, the second part of the pair \((...)\) does not matter, because it is never evaluated due to our not performing any updates in this example.

Now, we can evaluate the filter \(\varphi\) by \(\operatorname{eval}\, \left[\!\left[\varphi\right]\!\right]\, v = (\lambda \kappa. \operatorname{run}\, \left[\!\left[\varphi\right]\!\right])\, \operatorname{zero}\, v\). Because \(\varphi\) does not contain any labels, \(\left[\!\left[\varphi\right]\!\right]\) does not make any reference to \(\kappa\), therefore \(\operatorname{eval}\, \left[\!\left[\varphi\right]\!\right]\, v\) is equivalent to: \allowdisplaybreaks \begin{align*}
  \operatorname{run}\, \left[\!\left[\varphi\right]\!\right]\, v
  &= (\lambda \mathtt{repeat}. \operatorname{run}\, \left[\!\left[\mathtt{repeat}\right]\!\right]\, v)\, (Y\, (\lambda \mathtt{repeat}. \rho)) \\
  &=^{\left[\!\left[\cdot\right]\!\right]} (\lambda \mathtt{repeat}. \operatorname{run}\, \mathtt{repeat}\, v)\, (Y\, (\lambda \mathtt{repeat}. \rho)) \\
  &=^\beta \operatorname{run}\, (Y\, (\lambda \mathtt{repeat}. \rho))\, v \\
  &=^Y \operatorname{run}\, ((\lambda \mathtt{repeat}. \rho)\, (Y\, (\lambda \mathtt{repeat}. \rho)))\, v \\
  &=^\rho \operatorname{run}\, ((\lambda \mathtt{repeat}. \operatorname{pair}\, (\lambda v. \langle\operatorname{ok}v\rangle + \operatorname{run}\, \mathtt{repeat}\, v)\, (...))\, (Y\, (\lambda \mathtt{repeat}. \rho)))\, v \\
  &=^\beta \operatorname{run}\, (\operatorname{pair}\, (\lambda v. \langle\operatorname{ok}v\rangle + \operatorname{run}\, (Y\, (\lambda \mathtt{repeat}. \rho))\, v)\, (...))\, v \\
  &=^\beta \langle\operatorname{ok}v\rangle + \operatorname{run}\, (Y\, (\lambda \mathtt{repeat}. \rho))\, v \\
  &= \langle\operatorname{ok}v\rangle + \operatorname{run}\, \left[\!\left[\varphi\right]\!\right]\, v.
  \end{align*} This shows that the evaluation of \(\varphi\) on any input \(v\) yields an infinite stream of ``\(\operatorname{ok}v\)'' results.

\end{example}

\begin{example}[Labels]\label{ex:labels}

Let us consider the filter \(\varphi \equiv \mathopen{}\mathrel{\mathtt{label}}\, \$x\mathclose{}\;|\;\mathopen{}\mathrel{\mathtt{break}}\, \$x\mathclose{}\). We have: \begin{align*}
  \operatorname{eval}\, \left[\!\left[\varphi\right]\!\right]\, v
  &= (\lambda \kappa. \operatorname{run}\left[\!\left[\mathopen{}\mathrel{\mathtt{label}}\, \$x\mathclose{}\;|\;\mathopen{}\mathrel{\mathtt{break}}\, \$x\mathclose{}\right]\!\right])\, \operatorname{zero}\, v \\
  &= (\lambda \kappa\, v. \operatorname{label}\, \kappa\, ((\lambda \$x\, \kappa. \operatorname{run}\, \left[\!\left[\mathopen{}\mathrel{\mathtt{break}}\, \$x\mathclose{}\right]\!\right]\, v)\, \kappa\, (\operatorname{succ}\, \kappa)))\, \operatorname{zero}\, v \\
  &= \operatorname{label}\, \operatorname{zero}\, ((\lambda \$x\, \kappa. \langle\mathopen{}\mathrel{\mathtt{break}}\, \$x\mathclose{}\rangle)\, \operatorname{zero}\, (\operatorname{succ}\, \operatorname{zero})) \\
  &= \operatorname{label}\, \operatorname{zero}\, \langle\operatorname{break}\, \operatorname{zero}\rangle \\
  &= \langle\rangle
  \end{align*} It is interesting to note that if instead of \(\mathopen{}\mathrel{\mathtt{break}}\, \$x\mathclose{}\), we would have used a more complex filter, e.g.~\(\mathopen{}\mathrel{\mathtt{label}}\, \$y\mathclose{}\;|\;...\), then \(\$y\) would have been substituted by \(\operatorname{succ}\, \operatorname{zero}\) (which we can already see in our large equation above). This mechanism reliably allows us to generate fresh labels and to differentiate for each \(\operatorname{break}\) the corresponding \(\operatorname{label}\) that handles it.

\end{example}

\subsection{Builtin filters}\label{sec:builtin}

\newcommand{\topath}{\operatorname{to\_path}}

Many named filters commonly available in jq, such as \(\mathtt{error}\) and \(\mathtt{path}(f)\), cannot be implemented by definition. That means that jq interpreters have to provide such filters natively. In our formal semantics, that amounts to substituting unbound occurrences of terms like \(\mathtt{error}^0\) and \(\mathtt{path}^1\) in the compiled lambda term by their underlying implementations: \begin{alignat*}{4}
\mathtt{error}^0:{}& \mathcal{F}&&\coloneqq &&\operatorname{pair}\, (\lambda v. \langle\operatorname{err}\, v\rangle&)\, &(\lambda \sigma\, v. \langle\operatorname{err}\, v\rangle) \\
\mathtt{path}^1:{}& \mathcal{F}\to \mathcal{F}&&\coloneqq \lambda f. &&\operatorname{pair}\, (\lambda v. \operatorname{run}\, f\, v \mathop{>\!\!\!>\!=}\nolimits\operatorname{to\_path}&)\, &(\lambda \sigma\, v. \langle\operatorname{err}\, ...\rangle)
\end{alignat*} In particular, \(\mathtt{path}(f)\) is an important filter that runs \(f\) and returns the path of each output as array, yielding an error if an output does not have an associated path: \[\operatorname{to\_path}\!: \mathcal{V}_{\mathcal{P}}\to \langle\mathcal{R}\, \mathcal{V}\rangle \coloneqq \lambda v. \langle\operatorname{snd}\, v\, (\lambda p. \operatorname{arr}\, (\operatorname{map}_{\langle\rangle}\, \operatorname{ok}\, p))\, (\operatorname{err}\, \dots)\rangle\] The inverse filter for \(\mathtt{path}(f)\) is \(\mathtt{getpath}(f)\): For every path produced by \(f\), it yields the value the path is pointing to. It can be implemented by definition as \[\mathtt{getpath}(f) \coloneqq f \mathrel{\mathtt{as}}\$p\;|\;\mathopen{}\mathrel{\mathtt{reduce}} \$p[] \mathrel{\mathtt{as}}\$i\;\; (.;\; .[\$i])\mathclose{}\] and is included in jq's standard library. We will use \(\mathtt{path}\) and \(\mathtt{getpath}\) in \autoref{sec:pathless-updates}.

\subsection{Folding}\label{sec:folding}

In this subsection, we will define the functions ``\(\operatorname{reduce }\)'' and ``\(\operatorname{foreach}\)'' which underlie the semantics for the folding operators \(\mathopen{}\mathrel{\mathtt{reduce}} f_x \mathrel{\mathtt{as}}\$x\;\; (.;\; f)\mathclose{}\) and \(\mathopen{}\mathrel{\mathtt{foreach}} f_x \mathrel{\mathtt{as}}\$x\;\; (.;\; f;\; g)\mathclose{}\).

First, let us look at what the evaluation of the various folding filters expands to. Assuming that the filter \(f_x\) evaluates to \(\langle x_0, ..., x_n\rangle\), then \(\mathtt{reduce}\) and \(\mathtt{foreach}\) expand to:\footnote{ \texttt{jq} implements restricted versions of these folding operators that consider only the last output of \(f\) for the next iteration. That means that in \texttt{jq}, \(\mathopen{}\mathrel{\mathtt{reduce}} f_x \mathrel{\mathtt{as}}\$x\;\; (.;\;            f )\mathclose{}\) is equivalent to \(\mathopen{}\mathrel{\mathtt{reduce}} f_x \mathrel{\mathtt{as}}\$x\;\; (.;\; \mathtt{last}(f))\mathclose{}\). Here, we assume that the filter \(\mathtt{last}(f)\) returns the last output of \(f\) if \(f\) yields any output, else nothing. Furthermore, \texttt{jq} does not preserve the paths of values output by folding operators; for example, \texttt{{[}1{]}\ \textbar{}\ path(reduce\ 0\ as\ \$x\ (.;\ .{[}\$x{]}))} yields an error in \texttt{jq}, whereas our semantics yield the output \texttt{0}.} \begin{alignat*}{2}
\mathopen{}\mathrel{\mathtt{reduce}} f_x \mathrel{\mathtt{as}}\$x\;\; (.;\; f   )\mathclose{} ={}& x_0 \mathrel{\mathtt{as}}\$x\;|\;f & \quad
\mathopen{}\mathrel{\mathtt{foreach}} f_x \mathrel{\mathtt{as}}\$x\;\; (.;\; f;\; g)\mathclose{} ={}& x_0 \mathrel{\mathtt{as}}\$x\;|\;f\;|\;g, ( \\
|\;\; & ... &
    & ... \\
|\;\; & x_n \mathrel{\mathtt{as}}\$x\;|\;f &
    & x_n \mathrel{\mathtt{as}}\$x\;|\;f\;|\;g, ( \\
    &     &
    & \mathtt{empty})...)
\end{alignat*} The evaluation of \texttt{reduce} and \texttt{foreach} has a very similar shape. It differs only in two aspects: ``\texttt{foreach}'' outputs intermediate values (composed with \texttt{g}), and ``\texttt{reduce}'' outputs a final value (suppressed by \texttt{empty} in \texttt{foreach}).

\newcommand{\foldf}{\operatorname{fold}}

We will now define a function ``\(\operatorname{fold}\)'', which holds the logic common to \texttt{reduce} and \texttt{foreach}: \begin{align*}
\operatorname{fold}&{}\!: (T \to U \to \langle\mathcal{R}\, U\rangle) \to (T \to U \to \langle\mathcal{R}\, U\rangle) \to (U \to \langle\mathcal{R}\, U\rangle) \to \langle\mathcal{R}\, T\rangle \to U \to \langle\mathcal{R}\, U\rangle \\
&\coloneqq \lambda f\, g\, n. Y_2\, (\lambda F\, l\, v. l\, (\lambda h\, t. f\, h\, v \mathop{>\!\!\!>\!=}\nolimits(\lambda y. g\, h\, y + F\, t\, y))\, (n\, v))
\end{align*} Here, \(T\) is the type of list elements that the fold iterates over, and \(U\) is the type of the accumulator. When we use this function later, both \(T\) and \(U\) will be instantiated with \(\mathcal{V}_{\mathcal{P}}\) --- we distinguish between \(T\) and \(U\) here to clarify the difference between list elements and the accumulator.

This function takes two functions \(f\) and \(g\) that both take a list element and an accumulator and return a list of accumulator results, and a function \(n\) (for the nil case) from a final accumulator \(x\) to a list of accumulator results. From that, it creates a recursive function that takes a list of results \(l\) and an initial accumulator \(v\) and returns a list of accumulator results. This function folds over the elements in \(l\), starting from the accumulator value \(v\). For every element \(h\) in \(l\), \(f\) is evaluated with \(h\) and the current accumulator value \(v\) as input. Every output \(y\) of \(f\) is output after passing through \(g\), then used as new accumulator value with the remaining list \(t\). If \(l\) is empty, then \(v\) is called a \emph{final} accumulator value and \(n\, v\) is returned.

We use two different functions for \(n\); the first returns just its input, corresponding to \(\mathtt{reduce}\) which returns a final value, and the second returns nothing, corresponding to \(\mathtt{foreach}\). Instantiating ``\(\operatorname{fold}\)'' with these two functions, we obtain the following: \begin{alignat*}{4}
\operatorname{reduce }&\coloneqq \lambda f.     && \operatorname{fold}\, f\, (\lambda h\, v. \langle\rangle)\, && (\lambda v. \langle\operatorname{ok}v &&\rangle) \\
\operatorname{foreach}&\coloneqq \lambda f\, g. && \operatorname{fold}\, f\, g\, && (\lambda v. \langle&&\rangle)
\end{alignat*} Here, ``\(\operatorname{reduce }\)'' and ``\(\operatorname{foreach}\)'' are the functions used in \autoref{tab:eval-semantics}. Their types are: \begin{alignat*}{2}
\operatorname{reduce }&{}\!: (T \to U \to \langle\mathcal{R}\, U\rangle) &&\to \langle\mathcal{R}\, T\rangle \to U \to \langle\mathcal{R}\, U\rangle \\
\operatorname{foreach}&{}\!: (T \to U \to \langle\mathcal{R}\, U\rangle) \to (T \to U \to \langle\mathcal{R}\, U\rangle) &&\to \langle\mathcal{R}\, T\rangle \to U \to \langle\mathcal{R}\, U\rangle
\end{alignat*}

\section{Update Semantics}\label{sec:updates}

In this section, we will discuss how to evaluate updates \(f \mathbin{|{=}}g\). First, we show the path-based update semantics used in most jq interpreters. Then, we introduce our alternative, pathless update semantics, which avoid many problems of path-based updates. In \autoref{sec:update-eval}, we will give evidence that pathless updates enable higher performance by omitting the construction of temporary path data.

\subsection{Path-based updates}\label{sec:jq-updates}

jq's update mechanism works with \emph{paths}. A path is a sequence of indices \(i_j\) that can be written as \(.[i_1]...[i_n]\). It refers to a value that can be retrieved by the filter ``\(.[i_1]\;|\;...\;|\;.[i_n]\)''. Note that ``\(.\)'' is a valid path, referring to the input value.

The update operation ``\(f \mathbin{|{=}}g\)'' attempts to first obtain the paths of all values returned by \(f\), then for each path, it replaces the value at the path by \(g\) applied to it. Note that \(f\) may not produce new values; it may only return parts of its input.

\begin{example}\label{ex:arr-update}

Consider the input value \([[1, 2], [3, 4]]\). We can retrieve the arrays \([1, 2]\) and \([3, 4]\) from the input with the filter ``\(.[]\)'', and we can retrieve the numbers 1, 2, 3, 4 from the input with the filter ``\(.[]\;|\;.[]\)''. To replace each number with its successor, we run ``\((.[]\;|\;.[]) \mathbin{|{=}}.+1\)'', obtaining \([[2, 3], [4, 5]]\). Internally, in jq, this first builds the paths \(.[0][0]\), \(.[0][1]\), \(.[1][0]\), \(.[1][1]\), then updates the value at each of these paths with \(g\).

\end{example}

This approach can yield surprising results when the execution of the filter \(g\) changes the input value in a way that the set of paths changes midway. In such cases, only the paths constructed from the initial input are considered. This can lead to paths pointing to the wrong data, paths pointing to non-existent data, and missing paths.

\begin{example}\label{ex:obj-update-arr}

Consider the input value \(\left\{\text{"a"} \mapsto \left\{\text{"b"} \mapsto 1\right\}\right\}\) and the filter \((.[], .[][]) \mathbin{|{=}}g\), where \(g\) is \([]\). Executing this filter in jq first builds the path \(.[\text{"a"}]\) stemming from ``\(.[]\)'', then \(.[\text{"a"}][\text{"b"}]\) stemming from ``\(.[][]\)''. Next, jq folds over the paths, using the input value as initial accumulator and updating the accumulator at each path with \(g\). The final output is thus the output of \((.[\text{"a"}] \mathbin{|{=}}g)\;|\;(.[\text{"a"}][\text{"b"}] \mathbin{|{=}}g)\). The output of the first step \(.[\text{"a"}] \mathbin{|{=}}g\) is \(\left\{\text{"a"} \mapsto []\right\}\). This value is the input to the second step \(.[\text{"a"}][\text{"b"}] \mathbin{|{=}}g\), which yields an error because we cannot index the array \([]\) at the path \(.[\text{"a"}]\) by \(.[\text{"b"}]\).

\end{example}

We can also have surprising behaviour that does not manifest any error.

\begin{example}\label{ex:obj-update-obj}

Consider the same input value and filter as in \autoref{ex:obj-update-arr}, but now with \(g\) set to \(\left\{\text{"c"}\!: 2\right\}\). The output of the first step \(.[\text{"a"}] \mathbin{|{=}}g\) is \(\left\{\text{"a"} \mapsto \left\{\text{"c"} \mapsto 2\right\}\right\}\). This value is the input to the second step \(.[\text{"a"}][\text{"b"}] \mathbin{|{=}}g\), which yields \(\left\{\text{"a"} \mapsto \left\{\text{"c"} \mapsto 2, \text{"b"} \mapsto \left\{\text{"c"} \mapsto 2\right\}\right\}\right\}\). Here, the remaining path (\(.[\text{"a"}][\text{"b"}]\)) pointed to data that was removed by the update on the first path, so this data gets reintroduced by the update. On the other hand, the data introduced by the first update step (at the path \(.[\text{"a"}][\text{"c"}]\)) is not part of the original path, so it is \emph{not} updated.

\end{example}

We found that we can interpret many update filters by simpler filters, yielding the same output as jq in most common cases, but avoiding the problems shown above. To see this, let us see what would happen if we would interpret \((f_1, f_2) \mathbin{|{=}}g\) as \((f_1 \mathbin{|{=}}g)\;|\;(f_2 \mathbin{|{=}}g)\). That way, the paths of \(f_2\) would point precisely to the data returned by \(f_1 \mathbin{|{=}}g\), thus avoiding the problems depicted by the examples above. In particular, with such an approach, \autoref{ex:obj-update-arr} would yield \(\left\{\text{"a"} \mapsto []\right\}\) instead of an error, and \autoref{ex:obj-update-obj} would yield \(\left\{\text{"a"} \mapsto \left\{\text{"c"} \mapsto \left\{\text{"c"} \mapsto 2\right\}\right\}\right\}\).

In the remainder of this section, we will show semantics that extend this idea to all update operations. The resulting update semantics can be understood to \emph{interleave} calls to \(f\) and \(g\). By doing so, these semantics can abandon the construction of paths altogether, which results in higher performance when evaluating updates.

\subsection{Properties of pathless updates}\label{sec:update-props}

\begin{longtable}[]{@{}ll@{}}
\caption{Properties of pathless update semantics. }\label{tab:update-props}\tabularnewline
\toprule\noalign{}
\(\varphi\) & \(\varphi \mathbin{|{=}}\sigma\) \\
\midrule\noalign{}
\endfirsthead
\toprule\noalign{}
\(\varphi\) & \(\varphi \mathbin{|{=}}\sigma\) \\
\midrule\noalign{}
\endhead
\bottomrule\noalign{}
\endlastfoot
\(\mathtt{empty}\) & \(.\) \\
\(.\) & \(\sigma\) \\
\(f\;|\;g\) & \(f \mathbin{|{=}}(g \mathbin{|{=}}\sigma)\) \\
\(f, g\) & \((f \mathbin{|{=}}\sigma)\;|\;(g \mathbin{|{=}}\sigma)\) \\
\(f \mathrel{\mathtt{as}}\$x\;|\;g\) & \(((x_1 \mathrel{\mathtt{as}}\$x\;|\;g) \mathbin{|{=}}\sigma)\;|\;\dots\;|\;((x_n \mathrel{\mathtt{as}}\$x\;|\;g) \mathbin{|{=}}\sigma)\) \\
\(\mathopen{}\mathrel{\mathtt{if}} \$x \mathrel{\mathtt{then}} f \mathrel{\mathtt{else}} g \mathrel{\mathtt{end}}\mathclose{}\) & \(\mathopen{}\mathrel{\mathtt{if}} \$x \mathrel{\mathtt{then}} f \mathbin{|{=}}\sigma \mathrel{\mathtt{else}} g \mathbin{|{=}}\sigma \mathrel{\mathtt{end}}\mathclose{}\) \\
\(f \mathbin{/\!/}g\) & \(\mathopen{}\mathrel{\mathtt{if}} \mathtt{first}(f \mathbin{/\!/}\mathtt{false}) \mathrel{\mathtt{then}} f \mathbin{|{=}}\sigma \mathrel{\mathtt{else}} g \mathbin{|{=}}\sigma \mathrel{\mathtt{end}}\mathclose{}\) \\
\end{longtable}

\autoref{tab:update-props} gives a few properties that we want to hold for updates \(\varphi \mathbin{|{=}}\sigma\). Let us discuss these for the different filters \(\varphi\):

\begin{itemize}
\tightlist
\item
  \(\mathtt{empty}\): Returns the input unchanged.
\item
  ``\(.\)'': Returns the output of the update filter \(\sigma\) applied to the current input. Note that while jq only returns at most one output of \(\sigma\), these semantics return an arbitrary number of outputs.
\item
  \(f\;|\;g\): Updates at \(f\) with the update of \(\sigma\) at \(g\). This allows us to interpret \((.[]\;|\;.[]) \mathbin{|{=}}\sigma\) in \autoref{ex:arr-update} by \(.[] \mathbin{|{=}}(.[] \mathbin{|{=}}\sigma)\), yielding the same output as in the example.
\item
  \(f, g\): Applies the update of \(\sigma\) at \(g\) to the output of the update of \(\sigma\) at \(f\). We have already seen this at the end of \autoref{sec:jq-updates}.
\item
  \(f \mathrel{\mathtt{as}}\$x\;|\;g\): Applies \(\sigma\) at \(g\) for every value yielded by \(f\) bound to \(\$x\). Here, we suppose that \(f\) yields the outputs \(\langle x_1, \dots, x_n\rangle\). Then this property can be derived from \((f \mathrel{\mathtt{as}}\$x\;|\;g) \equiv ((x_1 \mathrel{\mathtt{as}}\$x\;|\;g), \dots, (x_n \mathrel{\mathtt{as}}\$x\;|\;g))\) and the concatenation property above. \autoref{sec:folding} suggests that is essentially equivalent to \(\mathopen{}\mathrel{\mathtt{reduce}} f \mathrel{\mathtt{as}}\$x\;\; (.;\; g \mathbin{|{=}}\sigma)\mathclose{}\) --- with the exception that \(\$x\) is not accessible in \(\sigma\).
\item
  \(\mathopen{}\mathrel{\mathtt{if}} \$x \mathrel{\mathtt{then}} f \mathrel{\mathtt{else}} g \mathrel{\mathtt{end}}\mathclose{}\): Applies \(\sigma\) at \(f\) if \(\$x\) holds, else at \(g\).
\item
  \(f \mathbin{/\!/}g\): Applies \(\sigma\) at \(f\) if \(f\) yields some output whose boolean value (see \autoref{sec:values}) is not false, else applies \(\sigma\) at \(g\). Here, \(\mathtt{first}(f)\) is a filter that returns the first output of its argument \(f\) if there is one, else the empty list.
\end{itemize}

While \autoref{tab:update-props} allows us to define the behaviour of several filters by reducing them to more primitive filters, there are several filters \(\varphi\) which cannot be defined this way. We will therefore give the actual pathless update semantics of \(\varphi \mathbin{|{=}}\sigma\) in \autoref{sec:pathless-updates} by defining \(\operatorname{upd}\, \left[\!\left[\varphi\right]\!\right]\, \sigma\, v\), not by translating \(\varphi \mathbin{|{=}}\sigma\) to equivalent filters.

\subsection{Pathless updates}\label{sec:pathless-updates}

We will now give update semantics that define the output of \(\operatorname{run}\, \left[\!\left[f \mathbin{|{=}}g\right]\!\right]\, v\) as referred to in \autoref{sec:semantics}. We will first recall from \autoref{tab:eval-semantics}: \[\operatorname{run}\, \left[\!\left[f \mathbin{|{=}}g\right]\!\right]\, v \coloneqq \operatorname{upd}\, \left[\!\left[f\right]\!\right]\, \sigma\, v, \text{where }
  \sigma\!: \mathcal{V}\to \langle\mathcal{R}\, \mathcal{V}\rangle \coloneqq \operatorname{run}\, \left[\!\left[g\right]\!\right]\]

\begin{longtable}[]{@{}ll@{}}
\caption{Pathless update semantics. Here, \(\varphi\) is a filter and \(\sigma\!: \mathcal{V}\to \langle\mathcal{R}\, \mathcal{V}\rangle\) is a function from a value to a list of value results. }\label{tab:update-semantics}\tabularnewline
\toprule\noalign{}
\(\varphi\) & \(\operatorname{upd}\, \left[\!\left[\varphi\right]\!\right]\, \sigma\, v\) \\
\midrule\noalign{}
\endfirsthead
\toprule\noalign{}
\(\varphi\) & \(\operatorname{upd}\, \left[\!\left[\varphi\right]\!\right]\, \sigma\, v\) \\
\midrule\noalign{}
\endhead
\bottomrule\noalign{}
\endlastfoot
\(.\) & \(\sigma\, v\) \\
\(..\) & \(\operatorname{upd}\left[\!\left[\mathopen{}\mathrel{\mathtt{def}} r\!\!: (.[]?\;|\;r), .;\;\mathclose{} r\right]\!\right]\, \sigma\, v\) \\
\(f\;|\;g\) & \(\operatorname{upd}\, \left[\!\left[f\right]\!\right]\, (\operatorname{upd}\, \left[\!\left[g\right]\!\right]\, \sigma)\, v\) \\
\(f, g\) & \(\operatorname{upd}\, \left[\!\left[f\right]\!\right]\, \sigma\, v \mathop{>\!\!\!>\!=}\nolimits\operatorname{upd}\, \left[\!\left[g\right]\!\right]\, \sigma\) \\
\(f \mathbin{/\!/}g\) & \(\operatorname{upd}\, ((\operatorname{run}\, \left[\!\left[f\right]\!\right]\, v \mathop{>\!\!\!>\!=}\nolimits\operatorname{trues})\, (\lambda \_\, \_. \left[\!\left[f\right]\!\right])\, \left[\!\left[g\right]\!\right])\, \sigma\, v\) \\
\(.[p]^?\) & \(\langle v[p]^? \mathbin{|{=}}\sigma\rangle\) \\
\(f \mathrel{\mathtt{as}}\$x\;|\;g\) & \(\operatorname{reduce }\, (\lambda x. (\lambda \$x. \operatorname{upd}\, \left[\!\left[g\right]\!\right])\, x\, \sigma)\, (\operatorname{run}\, \left[\!\left[f\right]\!\right]\, v)\, v\) \\
\(\mathopen{}\mathrel{\mathtt{if}} \$x \mathrel{\mathtt{then}} f \mathrel{\mathtt{else}} g \mathrel{\mathtt{end}}\mathclose{}\) & \(\operatorname{upd}\, ((\operatorname{bool}\, \$x)\, \left[\!\left[f\right]\!\right]\, \left[\!\left[g\right]\!\right])\, \sigma\, v\) \\
\(\mathopen{}\mathrel{\mathtt{break}}\, \$x\mathclose{}\) & \(\langle\operatorname{break}\, \$x\rangle\) \\
\(\mathopen{}\mathrel{\mathtt{reduce}} f_x \mathrel{\mathtt{as}}\$x\;\; (.;\; f   )\mathclose{}\) & \(\operatorname{reduce }_{\mathbin{|{=}}}\, (\lambda \$x. \operatorname{upd}\, \left[\!\left[f\right]\!\right])\, \sigma\, (\operatorname{run}\, \left[\!\left[f_x\right]\!\right]\, v)\, v\) \\
\(\mathopen{}\mathrel{\mathtt{foreach}} f_x \mathrel{\mathtt{as}}\$x\;\; (.;\; f;\; g)\mathclose{}\) & \(\operatorname{foreach}_{\mathbin{|{=}}}\, (\lambda \$x. \operatorname{upd}\, \left[\!\left[f\right]\!\right])\, (\lambda \$x. \operatorname{upd}\, \left[\!\left[g\right]\!\right])\, \sigma\, (\operatorname{run}\, \left[\!\left[f_x\right]\!\right]\, v)\, v\) \\
\(\mathopen{}\mathrel{\mathtt{def}} x(x_1;\; ...;\; x_n)\!\!: f;\;\mathclose{} g\) & \((\lambda x. \operatorname{upd}\, \left[\!\left[g\right]\!\right])\, (Y_{n+1}\, (\lambda x\, x_1\, ...\, x_n. \left[\!\left[f\right]\!\right]))\, \sigma\, v\) \\
\(x(f_1;\; ...;\; f_n)\) & \(\operatorname{upd}\, (x\, \left[\!\left[f_1\right]\!\right]\, ...\, \left[\!\left[f_n\right]\!\right])\, \sigma\, v\) \\
all others & \(\langle\operatorname{err}\, ...\rangle\) \\
\end{longtable}

\autoref{tab:update-semantics} shows the definition of \(\operatorname{upd}\, \left[\!\left[\varphi\right]\!\right]\, \sigma\, v\). Several of the cases for \(\varphi\), like ``\(.\)'', ``\(f\;|\;g\)'', ``\(f, g\)'', and ``\(\mathopen{}\mathrel{\mathtt{if}} \$x \mathrel{\mathtt{then}} f \mathrel{\mathtt{else}} g \mathrel{\mathtt{end}}\mathclose{}\)'' are simply relatively straightforward consequences of the properties in \autoref{tab:update-props}. We discuss the remaining cases for \(\varphi\):

\begin{itemize}
\tightlist
\item
  ``\(..\)'': Update the input value and all recursively contained values. This is very similar to the evaluation of ``\(..\)'' in \autoref{tab:eval-semantics}, but there, \(r\) is defined as ``\(., (.[]?\;|\;r)\)'', whereas here, it is ``\((.[]?\;|\;r), .\)'' --- the arguments of the concatenation are swapped. This leads values to be updated before the values that contain them. We discuss the effect of this in \autoref{ex:rec-update}.
\item
  \(f \mathbin{/\!/}g\): Updates using \(f\) if \(f\) yields some non-false value, else updates using \(g\). Here, we first call \(f\) as a ``probe''. If it yields at least one output that is considered ``true'' (see \autoref{sec:semantics} for the definition of \(\operatorname{trues}\)), then we update at \(f\), else at \(g\). This filter is unusual because is the only kind where a subexpression may be both evaluated (\(\operatorname{run}\, \left[\!\left[f\right]\!\right]\, v\)) as well as updated with (\(\operatorname{upd}\, \left[\!\left[f\right]\!\right]\, \sigma\, v\)).
\item
  \(.[p]^?\): Applies \(\sigma\) to the value at position \(p\) in the input, using the update operators in \autoref{sec:values}.
\item
  \(f \mathrel{\mathtt{as}}\$x\;|\;g\): Folds over all outputs of \(f\), using the input value \(v\) as initial accumulator and updating the accumulator by \(g \mathbin{|{=}}\sigma\), where \(\$x\) is bound in \(g\) to the current output of \(f\). The definition of ``\(\operatorname{reduce }\)'' is given in \autoref{sec:folding}. As explained in \autoref{sec:update-props}, we must bind \(\$x\) only in \(g\), not in \(\sigma\). That is why we use \((\lambda x. (\lambda \$x. \operatorname{upd}\, \left[\!\left[g\right]\!\right])\, x\, \sigma)\) instead of the simpler \((\lambda \$x. \operatorname{upd}\, \left[\!\left[g\right]\!\right] \sigma)\) in the first argument of ``\(\operatorname{reduce }\)''.
\item
  \(\mathopen{}\phi x \mathrel{\mathtt{as}}\$x\;\; (.;\; f)\mathclose{}\): Folds \(f\) over the values returned by \(\$x\). We will discuss this in \autoref{sec:folding-update}.
\item
  \(\mathopen{}\mathrel{\mathtt{def}} x(x_1;\; ...;\; x_n)\!\!: f;\;\mathclose{} g\): Defines a filter. This is defined analogously to \autoref{tab:eval-semantics}.
\item
  \(x(f_1;\; ...;\; f_n)\), \(x\): Calls a filter. This is defined analogously to \autoref{tab:eval-semantics}.
\item
  All others: Many filters \(\varphi\), such as \(\$x\), \([f]\), and \(\{\}\), yield fresh values instead of references to input data. In such cases, we yield an error, just like jq.
\end{itemize}

Our update semantics support all kinds of filters \(\varphi\) that are supported by jq, except for \(\mathopen{}\mathrel{\mathtt{label}}\, \$x\mathclose{}\;|\;g\) and \(\mathopen{}\mathrel{\mathtt{try}} f \mathrel{\mathtt{catch}} g\mathclose{}\). However, we never encountered any of these filters on the left-hand side of an update in the wild so far, and no user of our semantics requested this yet, see \autoref{sec:jaq}.

\begin{example}[Update compilation]

Let us consider the jq filter \((.[] \mathbin{|{=}}(.+.))\). When given an array as input, this filter outputs a new array where each value in the input array is replaced by the output of \(.+.\) on the value. The filter \(.+.\) returns the sum of the input and the input, effectively doubling its input. For example, when given the input \([1, 2, 3]\), the output of \(\varphi\) is \([2, 4, 6]\). Before we can execute the jq filter, we have to lower it to IR, e.g.~to \(\varphi \equiv (.[] \mathbin{|{=}}(. \mathrel{\mathtt{as}}\$x\;|\;\$x + \$x))\).\footnote{ Note that the actual lowering \((.[] \mathbin{|{=}}(. \mathrel{\mathtt{as}}\$x\;|\;. \mathrel{\mathtt{as}}\$y\;|\;\$x + \$y))\) is a bit longer than the \(\varphi\) given here, but because the two are semantically equivalent, we continue with \(\varphi\).} Let us now trace the execution of the filter, namely \(\operatorname{eval}\, \left[\!\left[\varphi\right]\!\right]\, v = (\lambda \kappa. \operatorname{run}\, \left[\!\left[\varphi\right]\!\right])\, \operatorname{zero}\, v\). Because \(\varphi\) does not use any labels, this is equivalent to just \(\operatorname{run}\, \left[\!\left[\varphi\right]\!\right]\, v = \operatorname{upd}\, \left[\!\left[(.[])\right]\!\right]\, \sigma\, v = (v[] \mathbin{|{=}}\sigma)\). Here, we introduced \(\sigma \equiv \operatorname{run}\, \left[\!\left[. \mathrel{\mathtt{as}}\$x\;|\;\$x + \$x\right]\!\right]\). We can further expand \(\sigma\): \begin{align*} \sigma
  &= \lambda v. \operatorname{run}\, \left[\!\left[. \mathrel{\mathtt{as}}\$x\;|\;\$x + \$x\right]\!\right]\, v \\
  &= \lambda v. \operatorname{run}\, \left[\!\left[.\right]\!\right] \mathop{>\!\!\!>\!=}\nolimits(\lambda \$x. \operatorname{run}\, \left[\!\left[\$x + \$x\right]\!\right]\, v) \\
  &= \lambda v. \langle\operatorname{ok}v\rangle \mathop{>\!\!\!>\!=}\nolimits(\lambda \$x. \langle\$x + \$x\rangle) \\
  &= \lambda v. \langle v + v\rangle
  \end{align*} In summary, \(\operatorname{eval}\left[\!\left[\varphi \right]\!\right]v = (v[] \mathbin{|{=}}(\lambda v. \langle v + v\rangle))\).

\end{example}

\begin{example}[Recursive update]\label{ex:rec-update}

Consider the input \([1]\) and the filter \(.. \mathbin{|{=}}f\), where \(f\) is \(\{a\!: .\}\). This filter is supposed to replace all values \(v\) recursively contained in the input by the object \(\{a\!: v\}\).

First, let us consider how \emph{path-based} updates handle this. The filter ``\(..\)'' is equivalent to \((\mathopen{}\mathrel{\mathtt{def}} r\!\!: ., (.[]?\;|\;r);\;\mathclose{} r)\) and yields the paths ``\(.\)'' (pointing to \([1]\)) and ``\(.[0]\)'' (pointing to \(1\)). The path-based update thus returns the same as \((. \mathbin{|{=}}f)\;|\;(.[0] \mathbin{|{=}}f)\). However, the result of ``\(. \mathbin{|{=}}f\)'' is \(\{a\!: [1]\}\), for which the path ``\(.[0]\)'' is invalid --- the path-based update fails with an error. The problem here is that ``\(..\)'' yields values closer to the root before values closer to the leaves. We call this strategy ``root-first'', as opposed to ``leaf-first''.

Next, let us consider how pathless updates would evaluate \(.. \mathbin{|{=}}f\) if ``\(..\)'' was interpreted like above. This, again, performs updates root-first. Using \autoref{tab:update-props}, we can show that this is equivalent to \(\mathopen{}\mathrel{\mathtt{def}} r\!\!: f\;|\;(.[]? \mathbin{|{=}}r);\;\mathclose{} r\). In this filter, it depends on the outputs of \(f\) how often the update \(r\) is performed; for example, if \(f\) always yields a non-empty array or object (as is the case for our given \(f\)), then \(r\) does not terminate. This significantly diverges from the behaviour of path-based updates.

This is why we use a different way to interpret ``\(..\)'' on the left-hand side of updates, namely as \(\mathopen{}\mathrel{\mathtt{def}} r\!\!: (.[]?\;|\;r), .;\;\mathclose{} r\). This performs updates leaf-first. We can show that using this, \(.. \mathbin{|{=}}f\) is equivalent to \(\mathopen{}\mathrel{\mathtt{def}} r\!\!: (.[]? \mathbin{|{=}}r)\;|\;f;\;\mathclose{} r\). Here, the number of times \(r\) calls \(f\) does \emph{not} depend on the output of \(f\), but only on the original input given to \(r\). This matches the behaviour of path-based updates much more closely, yet it terminates with a valid result, namely \(\{a\!: [\{a\!: 1\}]\}\).

\end{example}

\begin{example}[The Curious Case of Alternation]

The semantics of \((f \mathbin{/\!/}g) \mathbin{|{=}}\sigma\) can be rather surprising: For the input \(\left\{\text{"a"} \mapsto \operatorname{true}\right\}\), the filter \((.[\text{"a"}] \mathbin{/\!/}.[\text{"b"}]) \mathbin{|{=}}1\) yields \(\left\{\text{"a"} \mapsto 1\right\}\). This is what we might expect, because the input has an entry for \(\text{"a"}\). Now let us evaluate the same filter on the input \(\left\{\text{"a"} \mapsto \operatorname{false}\right\}\), which yields \(\left\{\text{"a"} \mapsto \operatorname{false}, \text{"b"} \mapsto 1\right\}\). Here, while the input still has an entry for \(\text{"a"}\) like above, its boolean value is \emph{not} true, so \(.[\text{"b"}] \mathbin{|{=}}1\) is executed. In the same spirit, for the input \(\left\{\right\}\) the filter yields \(\left\{\text{"b"} \mapsto 1\right\}\), because \(.[\text{"a"}]\) yields \(\operatorname{null}\) for the input, which also has the boolean value \(\operatorname{false}\), therefore \(.[\text{"b"}] \mathbin{|{=}}1\) is executed.

For the input \(\left\{\right\}\), the filter \((\operatorname{false}\mathbin{/\!/}.[\text{"b"}]) \mathbin{|{=}}1\) yields \(\left\{\text{"b"} \mapsto 1\right\}\). This is remarkable insofar as \(\operatorname{false}\) is not a valid path expression because it returns a value that does not refer to any part of the original input, yet the filter does not return an error. This is because \(\operatorname{false}\) triggers \(.[\text{"b"}] \mathbin{|{=}}1\), so \(\operatorname{false}\) is never used as path expression. However, running the filter \((\operatorname{true}\mathbin{/\!/}.[\text{"b"}]) \mathbin{|{=}}1\) \emph{does} yield an error, because \(\operatorname{true}\) triggers \(\operatorname{true}\mathbin{|{=}}1\), and \(\operatorname{true}\) is not a valid path expression.

Finally, on the input \([]\), the filter \((.[] \mathbin{/\!/}\mathtt{error}) \mathbin{|{=}}1\) yields \(\langle\operatorname{err}\, []\rangle\). That is because \(.[]\) does not yield any value for the input, so \(\mathtt{error} \mathbin{|{=}}1\) is executed, which yields an error.

\end{example}

We can perform path-based updates with pathless updates: In jq interpreters with path-based updates, the filter \(f \mathbin{|{=}}g\) is equivalent to \(\mathtt{getpath}(\mathtt{path}(f)) \mathbin{|{=}}g\). Here, \(\mathtt{path}(f)\) yields the paths to outputs of \(f\), then \(\mathtt{getpath}\) retrieves the corresponding value-paths, see \autoref{sec:builtin}. By replacing \(f \mathbin{|{=}}g\) with \(\mathtt{getpath}(\mathtt{path}(f)) \mathbin{|{=}}g\), we can force a path-based update, regardless of whether the used jq interpreter uses pathless or path-based updates. We use this to evaluate the performance of path-based updates in our interpreter that uses pathless updates, see \autoref{sec:update-eval}.

\subsection{Folding}\label{sec:folding-update}

In \autoref{sec:folding}, we have seen how to evaluate folding filters of the shape \(\mathopen{}\mathrel{\mathtt{reduce}} x \mathrel{\mathtt{as}}\$x\;\; (.;\; f)\mathclose{}\) and \(\mathopen{}\mathrel{\mathtt{foreach}} x \mathrel{\mathtt{as}}\$x\;\; (.;\; f;\; g)\mathclose{}\). Here, we will define pathless update semantics for these filters. These update operations are \emph{not} supported in jq 1.8 and yield an error; however, we will show that they arise quite naturally from previous definitions.

Let us start with an example to understand folding on the left-hand side of an update.

\begin{example}\label{ex:folding-update}

Let \(v = [[[2], 1], 0]\) be our input value and \(\varphi\) be the filter \(\mathopen{}\phi (0, 0) \mathrel{\mathtt{as}}\$x\;\; (.;\; .[\$x])\mathclose{}\). The regular evaluation of \(\varphi\) with the input value as described in \autoref{sec:semantics} yields \[\operatorname{run}\, \left[\!\left[\varphi\right]\!\right]\, v = \begin{cases}
    \langle\phantom{[[2], 1],\,} [2]\rangle & \text{if } \phi= \mathtt{reduce} \\
    \langle         [[2], 1],    [2]\rangle & \text{if } \phi= \mathtt{foreach}
  \end{cases}\] When \(\phi= \mathtt{foreach}\), the paths corresponding to the output are \(.[0]\) and \(.[0][0]\), and when \(\phi= \mathtt{reduce}\), the paths are just \(.[0][0]\). Given that all outputs have corresponding paths, we can update over them. For example, taking \(. + [3]\) as filter \(\sigma\), we should obtain the output \[\operatorname{upd}\, \left[\!\left[\varphi\right]\!\right]\, (\operatorname{run}\, \left[\!\left[\sigma\right]\!\right])\, v = \begin{cases}
    \langle[[[2, 3], 1\phantom{, 3}], 0]\rangle & \text{if } \phi= \mathtt{reduce} \\
    \langle[[[2, 3], 1         , 3 ], 0]\rangle & \text{if } \phi= \mathtt{foreach}
  \end{cases}\]

\end{example}

First, note that for folding filters, the lowering in \autoref{tab:lowering} and the defining equations in \autoref{sec:folding} only make use of filters for which we have already introduced update semantics in \autoref{tab:update-semantics}. This should not be taken for granted; for example, we originally lowered \(\mathopen{}\phi f_x \mathrel{\mathtt{as}}\$x\;\; (f_y;\; f)\mathclose{}\) to \[\lfloor f_y\rfloor \mathrel{\mathtt{as}}\$y\;|\;\mathopen{}\phi \lfloor f_x\rfloor \mathrel{\mathtt{as}}\$x\;\; (\$y;\; \lfloor f\rfloor)\mathclose{}\] instead of the more complicated lowering found in \autoref{tab:lowering}, namely \[. \mathrel{\mathtt{as}}\$x'\;|\;\lfloor f_y\rfloor\;|\;\mathopen{}\phi \lfloor\$x'\;|\;f_x\rfloor \mathrel{\mathtt{as}}\$x\;\; (.;\; \lfloor f\rfloor)\mathclose{}.\] While both lowerings produce the same output for regular evaluation, we cannot use the original lowering for updates, because the defining equations for \(\mathopen{}\phi x \mathrel{\mathtt{as}}\$x\;\; (\$y;\; f)\mathclose{}\) would have the shape \(\$y\;|\;...\), which is undefined on the left-hand side of an update. However, the lowering in \autoref{tab:lowering} avoids this issue by not binding the output of \(f_y\) to a variable, so it can be used on the left-hand side of updates.

To obtain an intuition about how the update evaluation of a fold looks like, we can take \(\mathopen{}\phi x \mathrel{\mathtt{as}}\$x\;\; (.;\; f;\; g)\mathclose{} \mathbin{|{=}}\sigma\), substitute the left-hand side by the defining equations in \autoref{sec:folding} and expand everything using the properties in \autoref{sec:update-props}. This yields \begin{alignat*}{2}
\mathopen{}\mathrel{\mathtt{reduce}} x \mathrel{\mathtt{as}}\$x\;\; (.;\; f   )\mathclose{} \mathbin{|{=}}\sigma
={}& x_0 \mathrel{\mathtt{as}}\$x\;|\;(f \mathbin{|{=}}( \\
 & ... \\
 & x_n \mathrel{\mathtt{as}}\$x\;|\;(f \mathbin{|{=}}(  \\
 & \sigma))...)) \\
\mathopen{}\mathrel{\mathtt{foreach}} x \mathrel{\mathtt{as}}\$x\;\; (.;\; f;\; g)\mathclose{} \mathbin{|{=}}\sigma
={}& x_0 \mathrel{\mathtt{as}}\$x\;|\;(f \mathbin{|{=}}((g \mathbin{|{=}}\sigma)\;|\;\\
 & ... \\
 & x_n \mathrel{\mathtt{as}}\$x\;|\;(f \mathbin{|{=}}((g \mathbin{|{=}}\sigma)\;|\;\\
 & .))...)).
\end{alignat*}

\begin{example}

To see the effect of above equations, let us reconsider the input value and the filters from \autoref{ex:folding-update}. Using some liberty to write \(.[0]\) instead of \(0 \mathrel{\mathtt{as}}\$x\;|\;.[\$x]\), we have: \[\varphi \mathbin{|{=}}\sigma = \begin{cases}
    .[0] \mathbin{|{=}}\phantom{\sigma\;|\;(}.[0] \mathbin{|{=}}\sigma   & \text{if } \phi= \mathtt{reduce} \\
    .[0] \mathbin{|{=}}\sigma\;|\;( .[0] \mathbin{|{=}}\sigma)  & \text{if } \phi= \mathtt{foreach}
  \end{cases}\]

\end{example}

We will now formally define the functions used in \autoref{tab:update-semantics}. For this, we first introduce a function \(\operatorname{fold}_{\mathbin{|{=}}}\), as counterpart to the function \(\operatorname{fold}\) in \autoref{sec:folding}. \begin{align*}
\mathcal U\, T\, U&\coloneqq T \to (U \to \langle\mathcal{R}\, U\rangle) \to U \to \langle\mathcal{R}\, U\rangle \\
\operatorname{fold}_{\mathbin{|{=}}}&{}\!: \mathcal U\, T\, U \to (T \to U \to \langle\mathcal{R}\, U\rangle) \to (U \to \langle\mathcal{R}\, U\rangle) \to \langle\mathcal{R}\, T\rangle \to U \to \langle\mathcal{R}\, U\rangle \\
&\coloneqq \lambda f\, g\, n. Y_2\, (\lambda F\, l\, v. l\, (\lambda h\, t. f\, h\, (\lambda x. g\, h\, x \mathop{>\!\!\!>\!=}\nolimits F\, t)\, v)\, (n\, v))
\end{align*} Using this function, we can now define \begin{alignat*}{4}
\operatorname{reduce }_{\mathbin{|{=}}} &\coloneqq \lambda f\,     &&\sigma. \operatorname{fold}_{\mathbin{|{=}}}\, f\, (\lambda h\, v. \langle\operatorname{ok}v\rangle)\, && \sigma \\
\operatorname{foreach}_{\mathbin{|{=}}} &\coloneqq \lambda f\, g\, &&\sigma. \operatorname{fold}_{\mathbin{|{=}}}\, f\, (\lambda h\, v. g\, h\, \sigma\, v)\, && (\lambda v. \langle\operatorname{ok}v\rangle)
\end{alignat*} The types of the functions are: \begin{alignat*}{2}
\operatorname{reduce }_{\mathbin{|{=}}}&{}\!: \mathcal U\, T\, U                        &&\to (U \to \langle\mathcal{R}\, U\rangle) \to \langle\mathcal{R}\, T\rangle \to U \to \langle\mathcal{R}\, U\rangle \\
\operatorname{foreach}_{\mathbin{|{=}}}&{}\!: \mathcal U\, T\, U \to \mathcal U\, T\, U &&\to (U \to \langle\mathcal{R}\, U\rangle) \to \langle\mathcal{R}\, T\rangle \to U \to \langle\mathcal{R}\, U\rangle
\end{alignat*}

\section{Implementation}\label{sec:impl}

We implemented two interpreters for the jq language, namely ujq and jaq. ujq is a minimalistic research interpreter that demonstrates that our semantics can be implemented \emph{compactly}. jaq is a full-featured interpreter that demonstrates that our semantics can also be implemented \emph{efficiently}.

\subsection{ujq}\label{ujq}

ujq is a translation of this paper's contents to Haskell, aiming to provide an minimal executable version of this paper. In particular, ujq implements lowering of a jq filter to IR (\autoref{sec:ir}) and execution of an IR filter (\autoref{sec:semantics}). \autoref{sec:ujq} explains in more detail how ujq relates to our semantics.

\subsection{jaq}\label{sec:jaq}

jaq is a full-featured jq interpreter written in Rust. Like ujq, jaq implements the semantics in \autoref{sec:semantics}, however, it adds a number of optimisation techniques, such as tail-call optimisation (TCO). That makes it more difficult to establish by looking at the code that jaq implements our semantics. However, this can be tested experimentally by comparing outputs of ujq and jaq. Furthermore, jaq not only implements the core jq language, but also all syntactic constructs omitted from this specification, such as modules, as well as more than 90\% of the filters in jq's standard library. It demonstrates that the semantics can be implemented efficiently. To ensure compatibility with \texttt{jq}, jaq has a test suite of more than 500 jq programs which yield the same output for \texttt{jq} and jaq.

The other main jq interpreters, namely \texttt{jq} and gojq, both compile a jq program to a list of imperative instructions and execute it, whereas jaq compiles a jq program to an abstract syntax tree and interprets it.

jaq's public issue tracker has been instrumental to identify which parts of our semantics cause compatibility problems for jq users. This particularly matters for our pathless update semantics, where we diverge most notably from \texttt{jq}. Users reported three issues caused by pathless update semantics; these were all about the semantics of ``\texttt{..\ \textbar{}=\ g}''. As explained in \autoref{ex:rec-update}, our original semantics could cause infinite loops. The semantics shown in this paper address this issue. The fact that there were three independent reports of the same issue related to updates shows that users extensively use updates in jaq and report their issues with it. The fact that no other issues with updates were reported so far indicates that our pathless update semantics preserve compatibility for a vast majority of users. Most reported issues concerned the command-line interface and filters in the standard library. There are currently no open issues related to the core jq semantics described in this article.

\subsection{Evaluation}\label{sec:eval}

We evaluated the jq interpreters \texttt{jq}, gojq, and jaq\footnote{ The used versions are \texttt{jq} 1.8.1, gojq 0.12.18, and jaq revision 80a0dc0. The accompanying material for this article includes the benchmark data, the used version of jaq, as well as ujq.} on a number of programming language interpreters written in the jq language:

\begin{itemize}
\tightlist
\item
  \texttt{bf.jq} (35 lines of code) by the author of gojq is an interpreter for the Brainfuck language. We run it on \texttt{fib.bf}, a Brainfuck program that calculates the first 13 Fibonacci numbers.
\item
  \texttt{wsjq.jq} (552 LOC) by Thalia Archibald is an interpreter for the Whitespace language \citep{wsjq}. We run it on \texttt{hopper.ws}, a Whitespace program by Dmitry Belov that outputs a portrait of Grace Hopper.
\item
  \texttt{jqjq.jq} (2837 LOC) by Mattias Wadman is an interpreter for the jq language \citep{jqjq}. We run it on its whole test suite consisting of over 200 tests, where each test executes a jq program inside jqjq.
\end{itemize}

While these are relatively unusual jq programs due to their large size, they are very useful to test compatibility of jq interpreters, because they exploit many various language features. Furthermore, they do not require large input data sets, so performance results only depend on the jq interpreter, not on other components such as JSON parsing.

All evaluated jq interpreters, including jaq, yield the same output for all benchmarks. This shows that our semantics enable a high degree of compatibility with other jq interpreters.

\begin{figure}
\centering
\pgfplotscreateplotcyclelist{redblue}{
  {black},
  {blue,fill=blue!30!white,mark=none},
  {red,mark=none,pattern=dots,pattern color=red},
}

\begin{tikzpicture}
\begin{axis}[
  x=2cm,
  ylabel=Normalised time,
  width=6.8cm,
  ybar,
  xtick=data,
  xtick style={draw=none},
  ymin=0,
  legend style={
    at={(0.5,1.05)},
    anchor=south,
    legend columns=-1,
  },
  enlarge y limits={abs=8ex,upper},
  enlarge x limits=0.2,
  nodes near coords,
  every node near coord/.append style={rotate=90, anchor=west},
  point meta=explicit,
  xticklabels={
    bf-fib,
    {wsjq-\\hopper},
    {jqjq-\\test},
  },
  xticklabel style={align=center},
  cycle list name=redblue,
]

\def\rowmax{max(
  \thisrow{time-jaq}, max(
  \thisrow{time-jq},
  \thisrow{time-gojq}))
}

\newcommand{\normplot}[1]{%
  \addplot table[
    x=f,
    y    expr=\thisrow{#1} / \rowmax,
    meta expr=\thisrow{#1}
  ] {eval/interpreters.tsv};
}

\normplot{time-jaq}
\normplot{time-jq}
\normplot{time-gojq}
\legend{jaq,jq,gojq}

\end{axis}
\end{tikzpicture}
\caption{Time to run a benchmark in various jq interpreters, in milliseconds. Lower is better.}
\label{fig:interpreters}
\end{figure}

The benchmarks were run on a Linux system with an AMD Ryzen 5 5500U. The results are given in \autoref{fig:interpreters}. On ``bf-fib'', gojq is leading slightly before jaq. On ``wsjq-hopper'', jaq achieves the best performance, being about 60 times (!) as fast as gojq and 5 times as fast as \texttt{jq}. On ``jqjq-test'', gojq achieves the best performance, being about twice as fast as jaq.\footnote{ We only recently found the relatively high runtime of jaq on jqjq to be likely caused by a suboptimal handling of tail calls in jaq. We are confident to be able to provide better performance results for the final article.}

\subsection{Update performance}\label{sec:update-eval}

We now compare the performance of path-based updates (as used in \texttt{jq} and gojq) and pathless updates (as used in jaq). We evaluate on two different inputs:

\begin{itemize}
\tightlist
\item
  \texttt{{[}range(1000000){]}} (an array of the shape \texttt{{[}0,\ ...,\ 999999{]}})
\item
  \texttt{\{"a":\ {[}range(1000000){]}\}} (an object of the shape \texttt{\{"a":\ {[}0,\ ...,\ 999999{]}\}})
\end{itemize}

The second input allows us to measure the impact of updating nested data. For both inputs, we evaluate the runtime of different actions, in order to determine the cost of different kinds of updates:

\begin{itemize}
\tightlist
\item
  Construction: Only construct the input (identity function). This serves as baseline, because all actions include input construction.
\item
  Native update: Update using the built-in update operator \texttt{\textbar{}=}. On jaq, this uses pathless updates, whereas on jq, this uses path-based updates.
\item
  Manual update: Update without \texttt{\textbar{}=}.
\item
  Path-based update: Update with \texttt{\textbar{}=}, forcing the usage of paths as shown in \autoref{sec:pathless-updates}.
\end{itemize}

\begin{longtable}[]{@{}lll@{}}
\caption{Evaluated filter \texttt{f} depending on input and action. \label{tab:update-eval}}\tabularnewline
\toprule\noalign{}
Action & \texttt{{[}range(1000000){]}} & \texttt{\{"a":\ {[}range(1000000){]}\}} \\
\midrule\noalign{}
\endfirsthead
\toprule\noalign{}
Action & \texttt{{[}range(1000000){]}} & \texttt{\{"a":\ {[}range(1000000){]}\}} \\
\midrule\noalign{}
\endhead
\bottomrule\noalign{}
\endlastfoot
Construction & \texttt{.} & \texttt{.} \\
Native update & \texttt{.{[}{]}\ \textbar{}=\ .} & \texttt{.a{[}{]}\ \textbar{}=\ .} \\
Manual update & \texttt{{[}.{[}{]}\ \textbar{}\ .{]}} & \texttt{\{a:\ .a\ \textbar{}\ {[}.{[}{]}\ \textbar{}\ .{]}\}} \\
Path-based update & \texttt{getpath(path(.{[}{]}))\ \textbar{}=\ .} & \texttt{getpath(path(.a{[}{]}))\ \textbar{}=\ .} \\
\end{longtable}

The filters that correspond to each of these actions are given in \autoref{tab:update-eval}. Every update applies the identity function (\texttt{.}) to all elements of the array contained in its input, which means that they return output equal to the input.

We evaluate the runtime of input construction \texttt{i} and action \texttt{f} by running \texttt{\$JQ\ -n\ \textquotesingle{}i\ \textbar{}\ f\ \textbar{}\ empty}.\footnote{The usage of \texttt{empty} is necessary to avoid printing the output, which can be very costly performance-wise.} For example, to evaluate jaq's native update performance on array input, we measure the time taken by \texttt{jaq\ -n\ \textquotesingle{}{[}range(1000000){]}\ \textbar{}\ .{[}{]}\ \textbar{}=\ .\ \textbar{}\ empty\textquotesingle{}}.

\begin{figure}
\centering
\pgfplotscreateplotcyclelist{redblue}{
  {red,mark=none,pattern=dots,pattern color=red},
  {blue,fill=blue!30!white,mark=none},
}

\begin{tikzpicture}
\begin{groupplot}[
  group style={group size=2 by 1, horizontal sep=0.5cm, yticklabels at=edge left},
  y=1.5cm,
  xlabel=Runtime,
  x unit=ms,
  axis x line=bottom,
  width=6.8cm,
  xbar,
  yticklabel style={align=left},
  y dir=reverse,
  ytick=data,
  ytick style={draw=none},
  xmin=0,
  enlarge x limits={abs=8ex,upper},
  enlarge y limits=0.2,
  nodes near coords,
  yticklabels={
    Construction,
    Manual update,
    Native update,
    Path-based update,
  },
  cycle list name=redblue,
  reverse legend,
]

\nextgroupplot[title={\texttt{[range(1000000)]}}]
\addplot+ table[y=f, x=time-gojq] {eval/upd-arr.tsv};
\addplot+ table[y=f, x=time-jq  ] {eval/upd-arr.tsv};
\addplot+ table[y=f, x=time-jaq ] {eval/upd-arr.tsv};
\legend{gojq,jq,jaq}

\nextgroupplot[title={\texttt{\{"a":\ [range(1000000)]\}}}]
\addplot+ table[y=f, x=time-gojq] {eval/upd-obj.tsv};
\addplot+ table[y=f, x=time-jq  ] {eval/upd-obj.tsv};
\addplot+ table[y=f, x=time-jaq ] {eval/upd-obj.tsv};

\end{groupplot}
\end{tikzpicture}
\caption{Runtime to construct and update input. Lower is better.}
\label{fig:update}
\end{figure}

The results are given in \autoref{fig:update}. Consider the results for array input \texttt{{[}range(1000000){]}}: We can see that native update performance differs enormously between \texttt{jq} and jaq: When subtracting the time for input construction, \texttt{jq} takes about 47 times (!) as long for the update as jaq.\footnote{ \((2041 - 106) \div (114 - 73) = 47.20\)} We can also see that in \texttt{jq} and gojq, (path-based) native updates are significantly \emph{slower} than manual updates, whereas in jaq, (pathless) native updates are \emph{faster} than manual updates. Finally, we can see that when forcing path-based updates, the performance of jaq plummets, arriving at the same order of magnitude as \texttt{jq}'s and gojq's native updates.\footnote{When forcing path-based updates in \texttt{jq} and gojq via \texttt{getpath(path(...))\ \textbar{}=\ ...}, the performance also decreases compared to its (path-based) native updates. That is because in this scenario, \texttt{jq} and gojq evaluate each path twice, namely a first time via \texttt{\textbar{}=} and a second time via \texttt{getpath(path(...))}.} That indicates that \texttt{jq}'s and gojq's low update performance is caused by path-based updates.

Now, let us look at the results for object input \texttt{\{"a":\ {[}range(1000000){]}\}}, in order to study the performance of nested updates, namely updating all values of an array inside an object. First, we can observe that the performance of manual updates in \texttt{jq}, gojq, and jaq, as well as native update performance in jaq, remain stable. That means that these kinds of updates are not impacted by nesting. On the other hand, the performance of path-based updates clearly decreases.

To sum it up: Path-based updates are significantly slower than manually updating data without the \texttt{\textbar{}=} operator. Furthermore, path-based update performance is impacted negatively by nesting. However, updates can be made to achieve higher performance than manual updates, by using our pathless update semantics.

\section{Conclusion}\label{conclusion}

We have shown syntax and semantics of a large subset of the jq programming language.

We first defined a subset of jq's filter syntax and a simpler subset (IR), and gave a lowering from concrete jq syntax to IR. We then gave formal semantics based on IR, by specifying how to compile a jq program to a lambda term. We discussed two strategies to interpret updates, namely path-based and pathless. We showed in our evaluation that our semantics can be implemented compactly (ujq) and efficiently (jaq), and that pathless updates can yield significant performance gains over path-based updates.

\appendix

\section{JSON values}\label{sec:json}

A JSON value \(v\) has the shape \[v \coloneqq \operatorname{null}\mathbin{\quad\|\quad}\operatorname{false}\mathbin{\quad\|\quad}\operatorname{true}\mathbin{\quad\|\quad}n \mathbin{\quad\|\quad}s \mathbin{\quad\|\quad}[v_0, ..., v_n] \mathbin{\quad\|\quad}\left\{k_0 \mapsto v_0, ..., k_n \mapsto v_n\right\},\] where \(n\) is a number and \(s\) is a string. A value of the shape \([v_0, ..., v_n]\) is called an \emph{array} and a value of the shape \(\left\{k_0 \mapsto v_0, ..., k_n \mapsto v_n\right\}\) is an unordered map from \emph{keys} \(k\) to values that we call an \emph{object}.\footnote{ The JSON syntax uses \(\left\{k_0\!: v_0, ..., k_n\!: v_n\right\}\) instead of \(\left\{k_0 \mapsto v_0, ..., k_n \mapsto v_n\right\}\). However, in this text, we use the \(\left\{k_0\!: v_0, ..., k_n\!: v_n\right\}\) syntax to denote the \emph{construction} of objects, and use \(\left\{k_0 \mapsto v_0, ..., k_n \mapsto v_n\right\}\) syntax to denote actual objects.} In JSON, object keys are strings.

We write ``\(\operatorname{err}\, ...\)'' to denote ``\(\operatorname{err}\, e\)'' where we do not want to specify \(e\). (In actual jq interpreters, \(e\) is frequently an error message string.)

The functions to construct arrays and objects, as well as to retrieve the \emph{boolean value}, are as follows: \begin{alignat*}{3}
& \operatorname{arr}&:{}& \langle\mathcal{R}\, \mathcal{V}\rangle \to \mathcal{R}\, \mathcal{V}&& \coloneqq \lambda l. \operatorname{sum}\, (l \mathop{>\!\!\!>\!=}\nolimits(\lambda v. \langle\operatorname{ok}\, [v])\rangle))\, [] \\
& \operatorname{obj}_0&:{}& \mathcal{V}&& \coloneqq \left\{\right\} \\
& \operatorname{obj}_1&:{}& \mathcal{V}\to \mathcal{V}\to \mathcal{R}\, \mathcal{V}&& \coloneqq \lambda k\, v. \begin{cases}
  \operatorname{ok}\left\{k \mapsto v\right\} & \text{if $k$ is a string} \\
  \operatorname{err}... & \text{otherwise}
\end{cases} \\
& \operatorname{bool}&:{}& \mathcal{V}\to \mathbb{B}&& \coloneqq \lambda v. \begin{cases}
  \operatorname{false}& \text{if $v = \operatorname{null}$ or $v = \operatorname{false}$} \\
  \operatorname{true}& \text{otherwise}
\end{cases}
\end{alignat*} The function ``\(\operatorname{arr}\)'' transforms a list of value results into a single value result: It returns an array if all list elements are values, and the first exception in the list otherwise. It uses a helper function \(\operatorname{sum}\) that takes a list \(l\) and an accumulator \(v\). It returns the sum of the accumulator and the list elements if they are all OK, otherwise it returns the first exception in the list. This uses the addition operator \(+\!: \mathcal{V}\to \mathcal{V}\to \mathcal{R}\, \mathcal{V}\). \[\operatorname{sum}\!: \langle\mathcal{R}\, \mathcal{V}\rangle \to \mathcal{V}\to \mathcal{R}\, \mathcal{V}\coloneqq \lambda l\, v. l\, (\lambda h\, t. h \mathop{>\!\!\!>\!=}\nolimits_{\mathcal{R}}(\lambda o. (v + o \mathop{>\!\!\!>\!=}\nolimits_{\mathcal{R}}\operatorname{sum}\, t)))\, (\operatorname{ok}\, v)\]

We are now going to define the access operators \(v[p]\) and the update operators \(v[p] \mathbin{|{=}}f\). For all these operators, it holds that if \(v\) is a boolean or a number, then the operator yields an error. Given that the definition of these operators differs greatly between different jq interpreters, we are only going to cover a few basic cases for these operators that all jq interpreters agree on. We omit all remaining cases, such as indexing with non-integer numbers, indexing with non-existing keys, and slicing (\(v[l:h]\)).

\newcommand{\appvp}{\operatorname{app\_vp}}

The access operator \(v[p]\) of a value-path \(v\) at position \(p\) is defined as follows: \[v[] \coloneqq \begin{cases}
  \sum_i v[i] & \text{if $\operatorname{fst}v = [v_0, ..., v_n]$} \\
  \sum_i v[k_i] & \text{if $\operatorname{fst}v = \left\{k_0 \mapsto v_0, ..., k_n \mapsto v_n\right\}$}
\end{cases}\] \[v[i] \coloneqq \begin{cases}
  \langle\operatorname{ok}\, (\operatorname{app\_vp}\, v\, v_i\, i)\rangle & \text{if $\operatorname{fst}v = [v_0, ..., v_n]$, $i \in \mathbb N$, and $i \leq n$} \\
  \langle\operatorname{ok}\, (\operatorname{app\_vp}\, v\, v_j\, i)\rangle & \text{if $\operatorname{fst}v = \left\{k_0 \mapsto v_0, ..., k_n \mapsto v_n\right\}$ and $k_j = i$}
\end{cases}\] Here, we use \(\sum_i v[i]\) to abbreviate \(v[0] + \dots + v[n]\). If there are no elements \(i\) to iterate over (i.e.~because the input value is an empty array or object), the sum returns the empty list \(\langle\rangle\). Furthermore, we use a helper function \(\operatorname{app\_vp}\) that takes a value-path \(v_p\), replaces its value by \(v\), and appends \(i\) to its path if present: \[\operatorname{app\_vp}\!: \mathcal{V}_{\mathcal{P}}\to \mathcal{V}\to \mathcal{V}\to \mathcal{V}_{\mathcal{P}}\coloneqq \lambda v_p\, v\, i.
\operatorname{pair}\, v\, (\operatorname{snd}\, v_p\, (\lambda p. \operatorname{some}\, (p + \langle i\rangle))\, \operatorname{none})\]

The update operator \(v[p] \mathbin{|{=}}f\) is defined as follows: \[v[] \mathbin{|{=}}f \coloneqq \begin{cases}
  \operatorname{arr}\, (\sum_i (f\, v_i))
  & \text{if } v = [v_0, ..., v_n] \\
  \operatorname{sum}(\sum_i \langle\operatorname{obj}_?\, k_i\, (f\, v_i)\rangle) \operatorname{obj}_0
  & \text{if $v = \left\{k_0 \mapsto v_0, ..., k_n \mapsto v_n\right\}$}
\end{cases}\] \[v[i] \mathbin{|{=}}f \coloneqq \begin{cases}
  \begin{alignedat}{3}
  \operatorname{arr}\, (&\textstyle\sum_{j=0}^{i-1}&&\langle\operatorname{ok}\, v_j\rangle + (f\, v_i\, (\lambda h\, t. \langle h\rangle)\, \langle\rangle) \\
  +{} &\textstyle\sum_{j=i+1}^n&&\langle\operatorname{ok}\, v_j\rangle)
  \end{alignedat}
  &
  \begin{aligned}
  & \text{if $v = [v_0, ..., v_n]$, $i \in \mathbb N$,} \\
  & \quad \text{and } i \leq n
  \end{aligned}
  \\
  \begin{alignedat}{3}
  \operatorname{sum}(&\textstyle\sum_{j=0}^{i-1}&&\langle\operatorname{obj}_1\, k_j\, v_j\rangle + \langle\operatorname{obj}_?\, k\, (f\, v_i)\rangle \\
  +{} &\textstyle\sum_{j=i+1}^n &&\langle\operatorname{obj}_1\, k_j\, v_j\rangle) \operatorname{obj}_0
  \end{alignedat}
  &
  \begin{aligned}
  & \text{if } v = \left\{k_0 \mapsto v_0, ..., k_n \mapsto v_n\right\} \\
  & \quad \text{and } k_j = i
  \end{aligned}
\end{cases}\] Here, we use a helper function for the case that \(v\) is an object. This function attempts to construct an object from a key and (the first element of) a list: \[\operatorname{obj}_?\!: \mathcal{V}\to \langle\mathcal{R}\mathcal{V}\rangle \to \mathcal{R}\mathcal{V}\coloneqq \lambda k\, l. l\, (\lambda h\, t. h \mathop{>\!\!\!>\!=}\nolimits_{\mathcal{R}}\lambda o. \operatorname{obj}_1\, k\, o)\, (\operatorname{ok}\operatorname{obj}_0)\]

\begin{example}

Let \(v = \operatorname{pair}\, [1, 2]\, \operatorname{none}\), i.e.~a value-path of an array without a path. That means that we do not know where the value comes from. Then we have that \(v[0] = \langle\operatorname{ok}\, (\operatorname{pair}\, 1\, \operatorname{none})\rangle\), \(v[1] = \langle\operatorname{ok}\, (\operatorname{pair}\, 2\, \operatorname{none})\rangle\), and \(v[] = v[0] + v[1]\).

We can update the value e.g.~with \(f \coloneqq \lambda x. \langle x + 1\rangle\). Then we have that \(v[0] \mathbin{|{=}}f = \operatorname{ok}\, [2, 2]\), \(v[1] \mathbin{|{=}}f = \operatorname{ok}\, [1, 3]\), and \(v[] \mathbin{|{=}}f = \operatorname{ok}\, [2, 3]\). Note that here, implicit conversion (\autoref{sec:implicit-conversion}) casts the value-path \(v\) to a value, allowing us to perform the update.

\end{example}

\begin{example}

Let \(v = \operatorname{pair}\, \left\{\text{"a"} \mapsto [1, 2]\right\}\, (\operatorname{some}\, \langle\rangle)\), i.e.~a value-path with an empty path. We have that \(v[\text{"a"}] = \langle\operatorname{ok}\, v'\rangle\), where \(v' = \operatorname{pair}\, [1, 2]\, (\operatorname{some}\, \langle\text{"a"}\rangle)\). This value-path \(v'\) records where its value \([1, 2]\) came from, namely from \(\text{"a"}\).

Accessing \(v'\) appends to its path, i.e. \(v'[0] = \langle\operatorname{ok}\, (\operatorname{pair}\, 1\, (\operatorname{some}\, \langle\text{"a"}, 0\rangle))\rangle\).

\end{example}

\section{Wellformedness}\label{sec:wf}

To be able to run a filter, we wish that its compiled form is a closed lambda term. We will now show under which conditions this is the case.

\newcommand{\ub}{\operatorname{ub}}

Informally, we say that a filter is \emph{wellformed} if all references to named filters, variables, and labels were previously bound. For example, the filter ``\(a, \$x\)'' is not wellformed because neither \(a\) nor \(\$x\) was previously bound, but the filter ``\(\mathopen{}\mathrel{\mathtt{def}} a\!\!: 1;\;\mathclose{} 2 \mathrel{\mathtt{as}}\$x\;|\;a, \$x\)'' is wellformed.

\autoref{tab:ub} specifies the unbound symbols in a filter. For this, it uses a set of symbol names \(c\) that may contain variables \(\$v_x\), labels \(\$l_x\), and filters \(x^n\). This relies on the renaming of symbols shown in \autoref{sec:hygiene}.

\begin{longtable}[]{@{}ll@{}}
\caption{Unbound symbols of an IR filter \(\varphi\) with respect to a context \(c\). }\label{tab:ub}\tabularnewline
\toprule\noalign{}
\(\varphi\) & \(\operatorname{ub}\, \varphi\, c\) \\
\midrule\noalign{}
\endfirsthead
\toprule\noalign{}
\(\varphi\) & \(\operatorname{ub}\, \varphi\, c\) \\
\midrule\noalign{}
\endhead
\bottomrule\noalign{}
\endlastfoot
\(n\), \(s\), \(.\), \(..\), \([]\), or \(\{\}\) & \(\emptyset\) \\
\(\$x\), \(-\$x\), or \(\mathopen{}\mathrel{\mathtt{break}}\, \$x\mathclose{}\) & \(\{\$x\} \setminus c\) \\
\([f]\) & \(\operatorname{ub}\, f\, c\) \\
\(\{\$x\!: \$y\}\) or \(\$x \mathrel{\circ}\$y\) & \(\operatorname{ub}\, \$x\, c \cup \operatorname{ub}\, \$y\, c\) \\
\(f \star g\) or \(\mathopen{}\mathrel{\mathtt{try}} f \mathrel{\mathtt{catch}} g\mathclose{}\) & \(\operatorname{ub}\, f\, c \cup \operatorname{ub}\, g\, c\) \\
\(.[p]^?\) & \(\bigcup_{\$x \in p} \operatorname{ub}\, \$x\, c\) \\
\(f \mathrel{\mathtt{as}}\$x\;|\;g\) & \(\operatorname{ub}\, f\, c \cup \operatorname{ub}\, g\, (c \cup \{\$x\})\) \\
\(\mathopen{}\mathrel{\mathtt{label}}\, \$x\mathclose{}\;|\;f\) & \(\operatorname{ub}\, f\, (c \cup \{\$x\})\) \\
\(\mathopen{}\mathrel{\mathtt{if}} \$x \mathrel{\mathtt{then}} f \mathrel{\mathtt{else}} g \mathrel{\mathtt{end}}\mathclose{}\) & \(\operatorname{ub}\, \$x\, c \cup \operatorname{ub}\, f\, c \cup \operatorname{ub}\, g\, c\) \\
\(\mathopen{}\phi f_x \mathrel{\mathtt{as}}\$x\;\; (.;\; f;\; g)\mathclose{}\) & \(\operatorname{ub}\, f_x\, c \cup \operatorname{ub}\, f\, (c \cup \{\$x\}) \cup \operatorname{ub}\, g\, (c \cup \{\$x\})\) \\
\(\mathopen{}\mathrel{\mathtt{def}} x^n(x_1;\; \dots;\; x_n)\!\!: f;\;\mathclose{} g\) & \(\operatorname{ub}\, f\, (c \cup \{x^n\} \cup \bigcup_i \{x_i\}) \cup \operatorname{ub}\, g\, (c \cup \{x^n\})\) \\
\(x^n(f_1;\; \dots;\; f_n)\) & \(\{x^n\} \setminus c \cup \bigcup_i \operatorname{ub}\, f_i\, c\) \\
\end{longtable}

When we have a set of builtin filters such as \(c = \{\mathtt{error}^0, \mathtt{path}^1\}\) (see \autoref{sec:builtin}), then a compiled filter \(\varphi\) is wellformed with respect to these filters iff \(\operatorname{ub}\, \varphi\, c = \emptyset\).

\section{Comparison between ujq and our semantics}\label{sec:ujq}

ujq implements a value type that is quite close to JSON, as well as the operations given in \autoref{sec:values}. However, the operations implemented in ujq are quite simplified and do not support all functionalities of their counterparts in \texttt{jq}; for example, \texttt{jq} supports the division of two strings, whereas ujq yields a runtime error in that case. Implementing all value operations like in \texttt{jq} would add a lot of code to ujq, without deepening our understanding of the core semantics. The feasibility of implementing more complete value operations within the framework established in \autoref{sec:values} is demonstrated by jaq.

There is one main difference between ujq and the semantics in this paper: In our semantics, compilation transforms an IR filter into a lambda term that can be readily executed. This compilation involves the construction of lambda terms with unbound variables, which relies on wellformedness (\autoref{sec:wf}) to ensure that the final result is a closed term. However, in Haskell, we cannot construct functions with unbound variables. Therefore, ujq carries along a \emph{context} that holds the currently assigned values to variables, filters, and break labels.\footnote{ We have also considered using an explicit context in this paper instead, but we found that this complicates presentation, while adding little to comprehension.} Furthermore, ujq \emph{interprets} IR filters, which is equivalent to performing the compilation step in this paper whenever a part of a filter is executed.

Despite these differences, we can still relatively easily see the equivalence between our formal semantics and ujq. To this end, we show an excerpt of our evaluation semantics in \autoref{tab:eval-excerpt} and the corresponding implementation in ujq:

\begin{Shaded}
\begin{Highlighting}[]
\OtherTok{run ::} \DataTypeTok{Value}\NormalTok{ v }\OtherTok{=\textgreater{}} \DataTypeTok{Filter} \OtherTok{{-}\textgreater{}} \DataTypeTok{Ctx}\NormalTok{ v }\OtherTok{{-}\textgreater{}} \DataTypeTok{ValP}\NormalTok{ v }\OtherTok{{-}\textgreater{}}\NormalTok{ [}\DataTypeTok{ValPR}\NormalTok{ v]}
\NormalTok{run f c}\OperatorTok{@}\DataTypeTok{Ctx}\NormalTok{\{vars\} vp }\OtherTok{=} \KeywordTok{case}\NormalTok{ f }\KeywordTok{of}
  \DataTypeTok{Id} \OtherTok{{-}\textgreater{}}\NormalTok{ [ok vp]}
  \DataTypeTok{Var}\NormalTok{ x }\OtherTok{{-}\textgreater{}}\NormalTok{ [ok }\OperatorTok{$}\NormalTok{ newVal }\OperatorTok{$}\NormalTok{ vars }\OperatorTok{!}\NormalTok{ x]}
  \DataTypeTok{Concat}\NormalTok{ f g }\OtherTok{{-}\textgreater{}}\NormalTok{ run f c vp }\OperatorTok{++}\NormalTok{ run g c vp}
  \DataTypeTok{Compose}\NormalTok{ f g }\OtherTok{{-}\textgreater{}}\NormalTok{ run f c vp }\OtherTok{\textasciigrave{}bind\textasciigrave{}}\NormalTok{ run g c}
  \DataTypeTok{Bind}\NormalTok{ f x g }\OtherTok{{-}\textgreater{}}\NormalTok{ run f c vp }\OtherTok{\textasciigrave{}bind\textasciigrave{}}\NormalTok{ (\textbackslash{}y }\OtherTok{{-}\textgreater{}}\NormalTok{ run g (bindVar x (val y) c) vp)}
\end{Highlighting}
\end{Shaded}

\begin{longtable}[]{@{}ll@{}}
\caption{Evaluation semantics (excerpt of \autoref{tab:eval-semantics}). }\label{tab:eval-excerpt}\tabularnewline
\toprule\noalign{}
\(\varphi\) & \(\operatorname{run}\, \left[\!\left[\varphi\right]\!\right]\, v\) \\
\midrule\noalign{}
\endfirsthead
\toprule\noalign{}
\(\varphi\) & \(\operatorname{run}\, \left[\!\left[\varphi\right]\!\right]\, v\) \\
\midrule\noalign{}
\endhead
\bottomrule\noalign{}
\endlastfoot
\(.\) & \(\langle\operatorname{ok}v\rangle\) \\
\(\$x\) & \(\langle\operatorname{ok}\$x\rangle\) \\
\(f, g\) & \(\operatorname{run}\, \left[\!\left[f\right]\!\right]\, v + \operatorname{run}\, \left[\!\left[g\right]\!\right]\, v\) \\
\(f\;|\;g\) & \(\operatorname{run}\, \left[\!\left[f\right]\!\right]\, v \mathop{>\!\!\!>\!=}\nolimits\operatorname{run}\, \left[\!\left[g\right]\!\right]\) \\
\(f \mathrel{\mathtt{as}}\$x\;|\;g\) & \(\operatorname{run}\, \left[\!\left[f\right]\!\right]\, v \mathop{>\!\!\!>\!=}\nolimits(\lambda \$x. \operatorname{run}\, \left[\!\left[g\right]\!\right]\, v)\) \\
\end{longtable}

This code shows the evaluation of the filters \(.\) (identity), \texttt{\$x} (variable), \(f, g\) (concatenation), \(f\;|\;g\) (composition), and \(f \mathrel{\mathtt{as}}\$x\;|\;g\) (binding). As mentioned above, ujq's \texttt{run} expects an additional context \texttt{c}, which is used to obtain the value bound to a variable in the \texttt{Var} case (via \texttt{vars\ !\ x}) and to bind a variable to a value in the \texttt{Bind} case (via \texttt{bindVar}). Furthermore, ujq explicitly performs conversions between values and value-paths, which we do implicitly in the formal semantics, as explained in \autoref{sec:implicit-conversion}. The code performs these conversions with the functions \texttt{newVal} in the \texttt{Var} case (convert a value to a value-path) and \texttt{val} in the \texttt{Bind} case (convert a value-path to a path).

ujq implements the built-in filter \texttt{path/1}, to allow evaluating the paths returned by a filter.

\bibliography{literature.bib}

\end{document}